\pdfoutput=1 
\documentclass{SciPost}
\usepackage{graphicx,amssymb,amsthm,epsfig,amsbsy,amsfonts,mathrsfs}
\usepackage[english,francais]{babel}

\usepackage{bbm}

\usepackage{url,hyperref}  

\usepackage{cite}

\definecolor{Cblue}{HTML}{045FB4}
\definecolor{Cred}{HTML}{DF0101}
\definecolor{CDarkred}{HTML}{8A0808}
\definecolor{Violet}{HTML}{6A0888}

\hypersetup{
    colorlinks=true, 
    linktoc=all,     
    linkcolor=CDarkred,
    citecolor=Cblue
}     

\renewcommand{\leq}{\leqslant}
\renewcommand{\geq}{\geqslant}

\newcommand{\ket}[1]{|\kern.3ex#1\kern.3ex\rangle}

\def\I{{\rm i}}                  
\def\dd{{\rm d}}                  

\newcommand{\abs}[1]{\ensuremath{\left| #1 \right|}}

\def\e{\mathrm{e}}

\def\kB{k_\mathrm{B}}

\newcommand{\moyQ}[1]{\ensuremath{\left\langle #1 \right\rangle}}

\newcommand{\moyT}[1]{\overline{#1}^{\mathrm{c}, \mathrm{g}}}

\newcommand{\moyTc}[1]{\overline{#1}^\mathrm{c}}
\newcommand{\moyTg}[1]{\overline{#1}^\mathrm{g}}

\def\Var{\mathrm{Var}}
\def\Cov{\mathrm{Cov}}

\def\O{\mathcal{O}}

\def\N{\mathcal{N}_+}

\def\L{\mathcal{L}}

\def\y{y}
\def\z{\xi}

\def\Fq{F_Q}
\def\Ft{F_T}

\def\ls{h}

\def\gc{\mathrm{g}}
\def\can{\mathrm{c}}

\def\Qs{_{\ket{ \{ n_k \}  }}}

\def\Ng{N}

\def\fug{\varphi}

\begin{document}

\selectlanguage{english}

\begin{center}
  {\Large \textbf{Fluctuations of observables for free
      fermions in a harmonic trap at finite temperature}}
\end{center}

\begin{center}
Aur\'elien Grabsch\textsuperscript{1*}, 
Satya N. Majumdar\textsuperscript{1},
Gr\'egory Schehr\textsuperscript{1},
Christophe Texier\textsuperscript{1}
\end{center}

\begin{center}
{\bf 1} LPTMS, CNRS, Univ. Paris-Sud, Universit\'e Paris-Saclay, 91405 Orsay, France
\\
* aurelien.grabsch@u-psud.fr
\end{center}

\begin{center}
\today
\end{center}


\section*{Abstract}

{\bf 
  We study a system of 1D noninteracting spinless fermions in a
  confining trap at finite temperature.  We first derive a useful and
  general relation for the fluctuations of the occupation numbers
  valid for arbitrary confining trap, as well as for both canonical
  and grand canonical ensembles.  Using this relation, we obtain
  compact expressions, in the case of the harmonic trap, for the
  variance of certain observables of the form of sums of a function of
  the fermions' positions, $\mathcal{L}=\sum_nh(x_n)$.
  Such observables are also called linear statistics of the positions.
  As anticipated, we demonstrate explicitly that these fluctuations do
  depend on the ensemble in the thermodynamic limit, as opposed to
  averaged quantities, which are ensemble independent.
  We have applied our general formalism to compute the fluctuations of
  the number of fermions $\mathcal{N}_+$ on the positive axis at
  finite temperature.  Our analytical results are compared to
  numerical simulations. We discuss the universality of the results
  with respect to the nature of the confinement.}

\vspace{10pt}
\noindent\rule{\textwidth}{1pt}
\tableofcontents\thispagestyle{fancy}
\noindent\rule{\textwidth}{1pt}
\vspace{10pt}

\hypersetup{
    linkcolor=Cred
}


\section{Introduction}
\label{sec:Introduction}

The recent experimental progresses in cold
atoms~\cite{BloDalZWe08,GioPitStr08,Ono16}, which have made accessible
new types of observables, have stimulated a renewed interest in the
study of fermionic systems over the past few years.  Although some
emphasis has been put on many body physics, many physical aspects of
the problem are captured by a simple non-interacting
picture. Moreover, there are practical ways to reach the
non-interacting regime
experimentally~\cite{BloDalZWe08,GioPitStr08,MarMakTur10}. Here, we
will restrict ourselves to this case, and consider a system of $N$
fermions in a one dimensional trap described by the Hamiltonian
\begin{equation}
  \hat{H} = \sum_{i=1}^N \left(
    \frac{\hat{p}_i^2}{2m} + V(\hat{x}_i)
  \right)
  \:.
\end{equation}
For specific choices of the confining potential $V(x)$, the positions
of the fermions at zero temperature can be mapped onto the eigenvalues
of random matrices. This relation is based on the fact that the ground
state of this system takes the form of a Slater determinant:
\begin{equation}
  \Psi_0(x_1,\ldots,x_N) = \frac{1}{\sqrt{N!}}
  \det[\psi_{i-1}(x_j)]_{1 \leq i,j \leq N}
  \:,
\end{equation}
where $\psi_k(x)$ is the one-body eigenfunction of energy
$\varepsilon_k$, with $k\in\mathbb{N}$.  For example, in the case of a
harmonic confinement $V(x)=\frac{1}{2} m \omega^2 x^2$, the joint
distribution of the positions, given by the modulus square of the
many-body wave function, reads:
\begin{equation}
  \abs{\Psi_0(x_1, \ldots,x_N)}^2
  = \frac{1}{\mathcal{Z}_N}
  \prod_{i<j} \abs{x_i-x_j}^2
  \prod_{i=1}^N \e^{-m \omega x_i^2/\hbar}
  \:,
\end{equation}
where $\mathcal{Z}_N$ is a normalisation constant. It is exactly the
distribution of the eigenvalues of matrices in the Gaussian unitary
ensemble (GUE)~\cite{Meh04,For10}. Similarly, in the case of an
infinite square well, the distribution of the positions can be mapped
onto the Jacobi unitary ensemble
(JUE)~\cite{CunMezOCo17,LacLDoMajSch17}. This connection to random
matrices has allowed to study many properties of the ground state,
like the density, the correlations, the number fluctuations and
entanglement
entropy~\cite{MarMajSchViv14a,CalLDoMaj15,DeLDoMajSch16,MarMajSchViv16}.
It is interesting to remark that in the cold atom literature~\cite{GleWonSch00,MinVigPat01,Cas07}, some
global results were derived using the local density approximation
(LDA), i.e. the Thomas-Fermi approximation, without realizing the
connection to RMT.

A remarkable recent achievement is the development of Fermi quantum
microscopes~\cite{CheNicOkaGerRamBakWasLomZwi15,HalHudKelCotPeaBruKuh15,ParHubMazChiSetWooBlaGre15},
allowing the direct measurement of the fermions' positions in a
confining trap.  Motivated by this context, several theoretical
studies have focused on different observables counting the fermions in
a given spatial domain (see
e.g. Refs.~\cite{Eis13,MarMajSchViv14a,CalLDoMaj15,MarMajSchViv16}).
One such observable is the number $\N$ of fermions on the positive
axis:
\begin{equation}
  \label{eq:DefNp}
  \N = \sum_{n=1}^N \Theta(x_n)
  \:,
  \quad
  \text{ with }
  \quad
  \Theta(x) = \left\lbrace
    \begin{array}{ll}
      1 & \text{ if } x > 0 \:, \\
      0 & \text{ otherwise.}
    \end{array}
  \right.
\end{equation}
Thanks to the mapping to the random matrix theory (RMT), the number
$\N$ of fermions in the domain $x>0$ at zero temperature is precisely
the number of positive eigenvalues of GUE random matrices.  In the RMT
literature, this number is known as the \textit{index}, and its
statistical properties have been studied for various RMT ensembles,
including the GUE \cite{MajNadScaViv09,MajNadScaViv11} and the Cauchy
ensemble~\cite{MarMajSchViv14}.
For GUE, the mean value of $\mathcal{N}_+$ is trivially $N/2$, but the variance is nontrivial, given by
\begin{equation}
  \label{eq:VarT0}
  \left. \Var(\N) \right|_{T=0} 
  \simeq
  \frac{1}{2\pi^2} \ln N
  + c
  \:,
  \qquad
  c = \frac{1+\gamma+3\ln 2}{2\pi^2}
  \:,
\end{equation}
where $\gamma$ the Euler-Mascheroni constant.  In the context of
fermions, this finite value characterizes the quantum
fluctuations. The logarithmic behaviour can be related to the
anticorrelation of fermions discussed below, see
Eq.~\eqref{eq:DensityDensityCorrelations}.  However in most
experiments, the measurements are done at low but finite temperature.
The zero temperature results obtained from RMT are therefore not
sufficient to address these finite temperature properties.
Our goal here will be to characterise the effect of thermal
fluctuations on the variance of $\N$ and to generalise the result
\eqref{eq:VarT0} to finite temperature.

Statistical physics provides several tools to analyse thermal
fluctuations.  When quantum correlations are dominant, like for the
problem we aim to study, it is well known that the most efficient
approach is supplied by the grand canonical ensemble in which the
temperature $T$ and the chemical potential $\mu$ are fixed, while the energy and the number of fermions fluctuate.
Many-body quantum eigenstates are conveniently labelled by a set of
\textit{occupation numbers} $\{n_k\}$, where $n_k=1$ if the individual
eigenstate $\psi_k(x)$ is occupied by one fermion and $n_k=0$
otherwise.  The grand canonical weight is:
\begin{equation}
  \label{eq:MeasureGC}
  \mathscr{P}_\gc(\{ n_k \}) =
  \frac{1}{Z_\gc } \: 
  \e^{-\beta \sum_k n_k (\varepsilon_k-\mu)} 
  \:,
\end{equation}
where 
\begin{equation}
  \label{eq:ParFctGC}
  Z_\gc(\fug) =
  \sum_{\{ n_k \}} \fug^{\sum_k n_k} \: \e^{-\beta \sum_k n_k
    \varepsilon_k}
\end{equation}
is the grand canonical partition function.  $\beta = 1/(k_B T)$ is the
inverse temperature and $\fug = \e^{\beta\mu}$ the fugacity.  In
atomic traps, the number of atoms is fixed, which is best described by
the microcanonical or canonical ensembles.  Furthermore, due to the
evaporative cooling techniques, the number of atoms is only moderately
large, $N\sim10^4$ to $10^7$, and the equivalence between statistical
physics ensembles is questionable.  This has recently motivated
several works where basic quantities, such as occupation numbers,
energy, specific heat, etc, have been analysed in the canonical and
microcanonical ensembles (see for instance
Refs.~\cite{SchMed96,Pol96,HolKalKir98,ChaMekZam99,Pra00,GreMajSch17}).
In the canonical ensemble, the number $N$ of fermions is fixed
(instead of the chemical potential) and the Gibbs weight is
\begin{equation}
  \label{eq:MeasureCan}
  \mathscr{P}_\can(\{ n_k \}) =
  \frac{1}{Z_\can}
  \:
  \e^{-\beta \sum_k n_k \varepsilon_k}
  \:
  \delta_{\sum_k n_k, N}
  \:,
\end{equation}
where  
\begin{equation}
  \label{eq:PartFctCan}
  Z_\can(N) = \sum_{\{ n_k \}} \e^{-\beta \sum_k n_k \varepsilon_k} 
  \: \delta_{\sum_k n_k, N}
\end{equation}
is the canonical partition function.  The constraint on the number of
particles included in the distribution however makes the calculations
much more difficult in practice.  For large $N$, deviations from the
thermodynamic limit, and thus differences between predictions from the ensembles, are
supposed to be small. However, it is worth stressing that the
equivalence of ensembles is valid only for thermodynamic quantities
(averaged observables), and not for their
fluctuations~\cite{PatBea11,TexRou17}, which are our main interest
here.  In the present article, we will introduce a general formalism
allowing to study the fluctuations of a wide class of observables of
the form
\begin{equation}
  \label{eq:defLinStat}
  \L = \sum_{n=1}^N \ls(x_n)
  \:,
\end{equation}
known as \textit{linear statistics} of the positions of the fermions,
where $\ls$ is any given function, not necessarily linear.  
For example, the potential energy in a harmonic trap corresponds to $\ls(x)=x^2$, whereas the index $\N$, under
consideration here, corresponds to $\ls(x) = \Theta(x)$.
Most theoretical studies of fluctuations at finite temperature were
performed in the grand canonical
ensemble~\cite{MosNeuSha94,GarVer03,GioPitStr08,DeLDoMajSch15,DeLDoMajSch16},
with the exception of Ref.~\cite{GreMajSch17}, where the specific form
$\sum_nx_n^2$ has allowed an exact calculation at all temperatures in
both ensembles (see also Ref.~\cite{LieWan17}).  Our aim here is to
introduce a general framework to analyse the fluctuations of arbitrary
linear statistics within both the grand canonical and canonical
ensembles.

\subsection{Summary of the main results}

The fluctuations of the linear statistics $\mathcal{L}=\sum_nh(x_n)$
have two origins at finite temperature.  For a given many-body quantum
state $\ket{\{n_k\}}$, labelled by the occupation numbers, the
positions $x_n$'s fluctuate due to quantum fluctuations.  In addition,
the occupation numbers $n_k$'s themselves fluctuate at finite
temperature --this is the thermal fluctuations. To characterize these
thermal fluctuations, we have found a general relation for the
correlator of occupation numbers:
\begin{equation}
  \label{eq:RelCovOccNb}
  \moyT{n_{k}n_{l}}
  = (\mp) \frac{
    \e^{\beta \varepsilon_k} \: \moyT{n_{k}}
    -\e^{\beta \varepsilon_l} \: \moyT{n_{l}}
  }{\e^{\beta \varepsilon_k}-\e^{\beta \varepsilon_l}}
  \:,
  \quad
  \left\lbrace
    \begin{array}{ll}
      - & \text{ for bosons,}\\
      + & \text{ for fermions,}
    \end{array}
  \right.
\end{equation}
where $\moyT{\cdots}$ denotes the thermal average (see also
\cite{GirGraTex17}).  We stress that this relation
\eqref{eq:RelCovOccNb} is a very general one, independent of the
confining trap.  Moreover it is valid both in the canonical
($\can$) and grand canonical ensemble
($\gc$). Note that in the grand canonical case, where the mean
occupation numbers are given respectively by the Bose-Einstein and
Fermi-Dirac distributions, this relation leads to:
$\moyTg{n_{k}n_{l}}=\moyTg{n_{k}} \: \moyTg{n_{l}}$.  This is the
well-known independence of energy levels.  This
relation~(\ref{eq:RelCovOccNb}) plays a crucial role for the
derivation of our subsequent results.

\begin{figure}[!ht]
\centering
\includegraphics[width=0.4\textwidth]{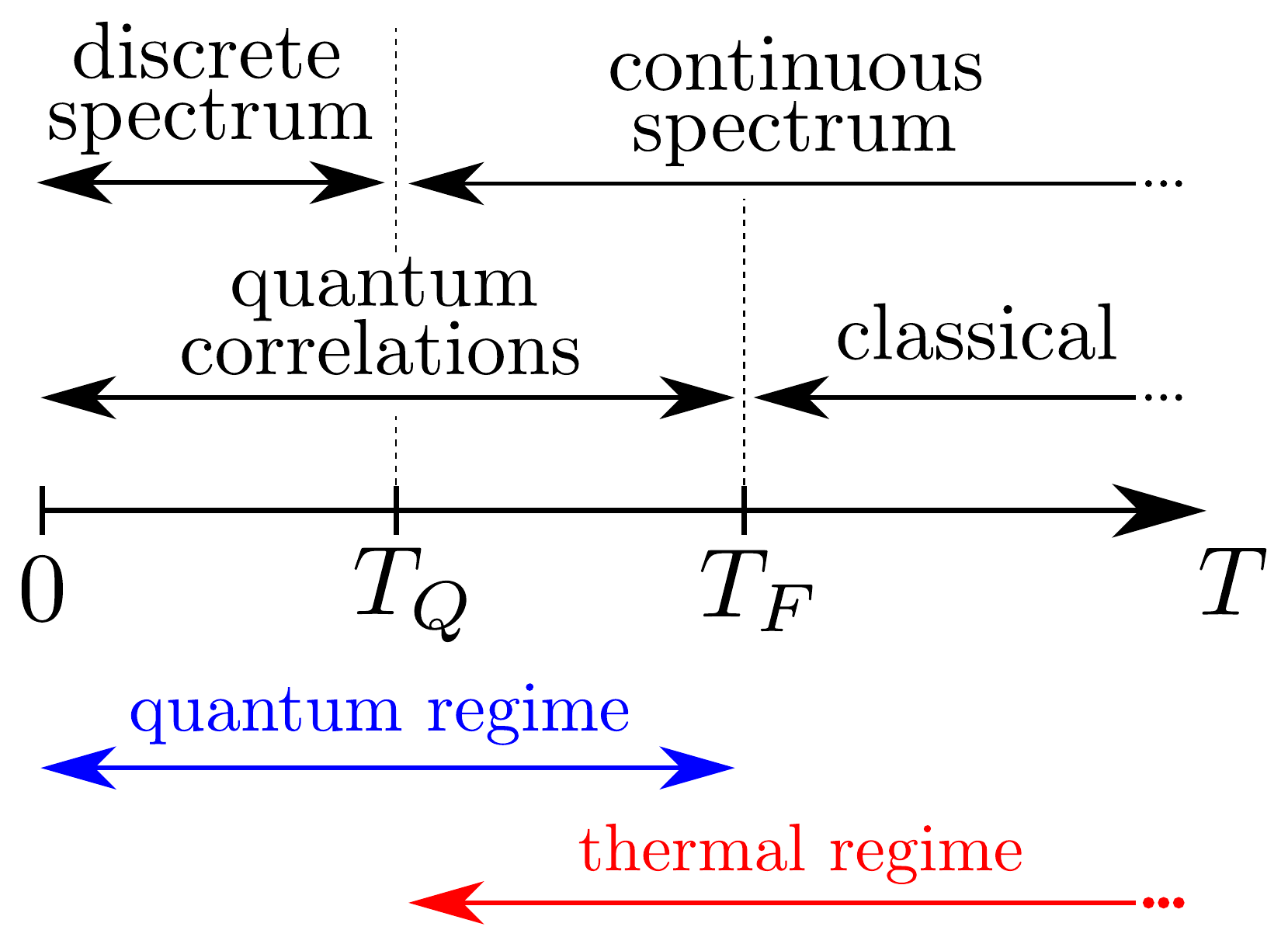}
\caption{\it Sketch of the different temperature regimes.}
\label{fig:Regimes}
\end{figure}

We have studied a 1D system of $N$ fermions in a harmonic trap $V(x) =
\frac{1}{2} m \omega^2 x^2$, where $N$ is either fixed (canonical
ensemble) or fluctuating (grand canonical ensemble).  We are
interested in the fluctuations of the number $\N$ of particles in the
domain $x>0$. We can already identify two temperature scales, which
will play a role below:
\begin{itemize}
\item a quantum scale $T_Q = \hbar \omega/k_B$: when $T\sim T_Q$, the
  small thermal energy allows only a few excitations near the Fermi
  level and the discreteness of the spectrum matters. We refer to this
  case as the \textit{quantum regime};
\item the Fermi temperature $T_F = N \hbar \omega/k_B$: 
  when $T\sim T_F$, the system is dominated by large
  thermal fluctuations. All energy levels contribute and the spectrum
  can be considered as continuous. Thus we call it the
  \textit{thermal regime}.
\end{itemize}
The quantum regime covers the transition between the regime where the
spectrum should be considered as discrete ($T\ll T_Q$) and the regime
where it can be considered as continuous ($T\gg T_Q$), while the
thermal regime describes the crossover between the regime dominated by
quantum fluctuations ($T\ll T_F$) and the classical regime ($T\gg
T_F$)~\cite{Tex15book}, cf.~Fig.\ref{fig:Regimes}. Note that, while
in the canonical ensemble $N$ is fixed, in the grand canonical
ensemble it is a fluctuating random variable.  Then, in the grand
canonical ensemble, the Fermi temperature is defined as $T_F =
\moyTg{N} \hbar \omega/k_B$.  Henceforth, we will denote the variance
of the total number of particles by $\Var_\gc(N)$ (the properties of
this variance, in particular its $T$-dependence, are discussed in
Appendix~\ref{app:VarN}; it is plotted in Fig.~\ref{fig:PltVarN} as a
function of temperature).  We denote $\Var(\N)$ the variance of $\N$
which is computed in these two different regimes and the different
statistical ensembles.

\begin{figure}[!ht]
  \centering
  \includegraphics[width=0.45\textwidth]{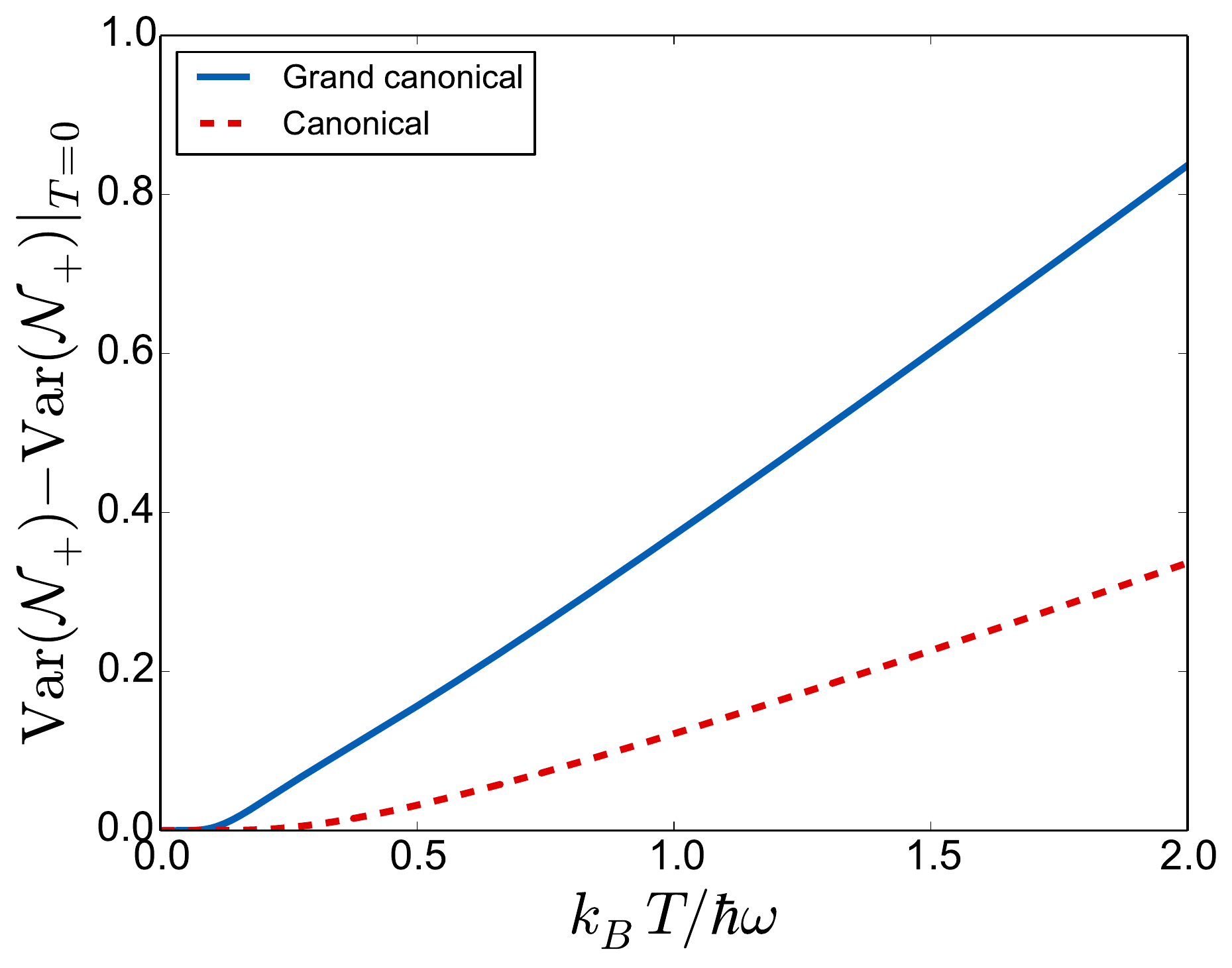}
  \includegraphics[width=0.45\textwidth]{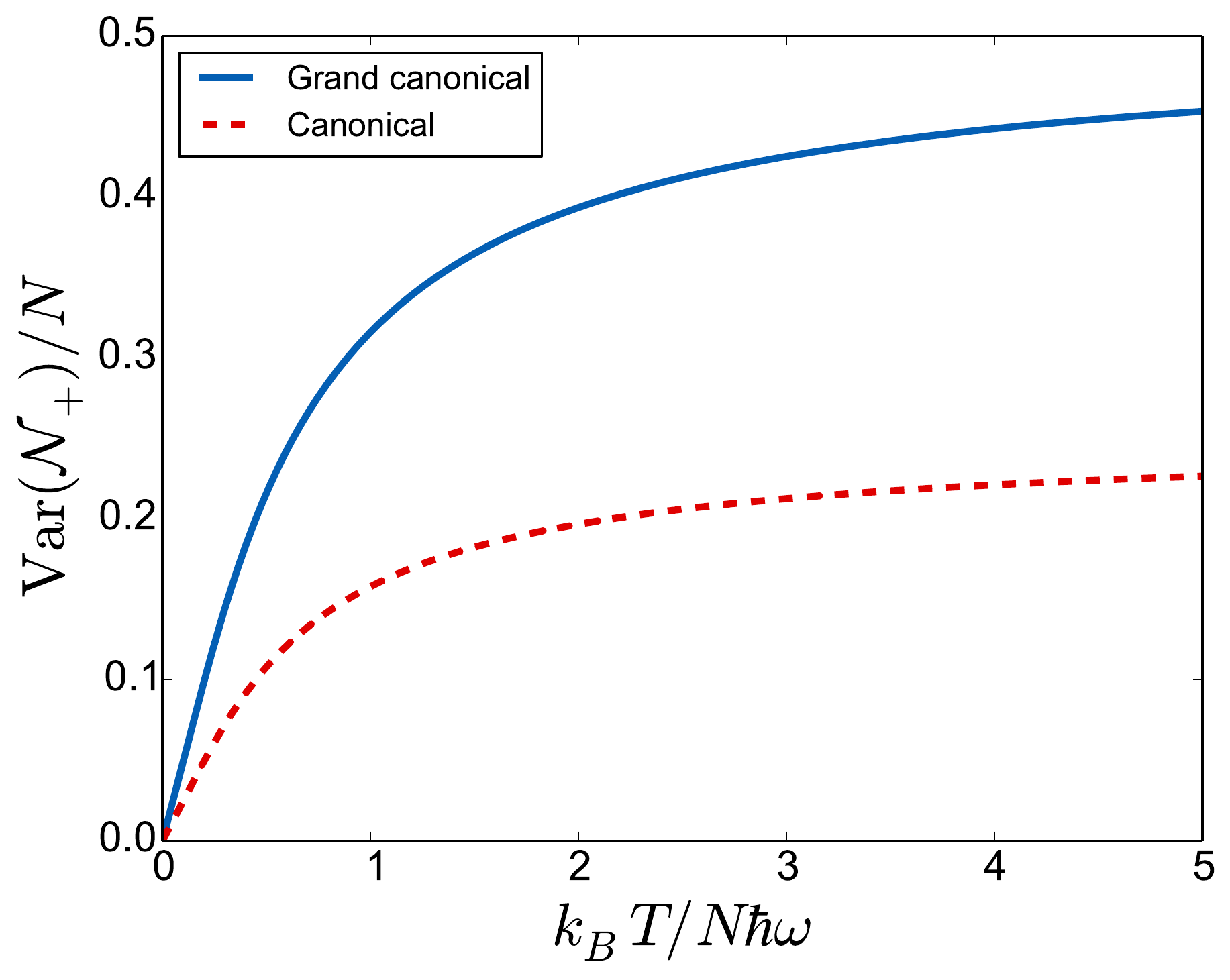}
  \caption{\textit{Variance of the number $\N$ of fermions in the domain $x>0$
    as a function of the temperature, both in the canonical and grand
    canonical ensembles. }
    Left:\textit{ quantum regime $T \sim T_Q$. }
    Right:
    \textit{thermal regime $T \sim T_F$. In the grand canonical case, the
    number of particles $N$ must be replaced by its mean value
    $\moyTg{N}$.}}
  \label{fig:PltVar}
\end{figure}

In the quantum regime $T \sim T_Q$, we have obtained:
\begin{equation}
  \Var(\N)
  \simeq
  \left\lbrace
    \begin{array}{ll}
      \displaystyle
      \left. \Var(\N) \right|_{T=0} +
      \Fq(\xi) 
      &
      \text{ canonical,}
      \\[0.3cm]
      \displaystyle
      \left. \Var(\N) \right|_{T=0} +
      \Fq(\xi) 
      +  \frac{1}{4} \: \Var_\gc(N)
      &
      \text{ grand canonical,}
    \end{array}
  \right.
\end{equation}
where the zero temperature variance $\left. \Var(\N) \right|_{T=0}$ is
given by Eq.~(\ref{eq:VarT0}), and we introduced
\begin{equation}
  \label{eq:FqUniv}
  \Fq \left(\z = \frac{T_Q}{T} \right)
  = \frac{2}{\pi^2} \sum_{n=1}^\infty
  \frac{1}{2n-1}
  \frac{1}{\e^{(2n-1) \z} - 1}
  \:.
\end{equation}
In the grand canonical case, the variance receives a contribution
proportional to the fluctuations of the total number of particles,
$\Var_\gc(\Ng)$.
This observation is also valid in the thermal regime $T \sim T_F$,
where our result reads:
\begin{equation}
  \Var(\N)
  \simeq
  \left\lbrace
    \begin{array}{ll}
      \displaystyle
      N \Ft(\y)
      &
      \text{ canonical,}
      \\[0.3cm]
      \displaystyle
      \moyTg{N} \Ft(\y) +
      \frac{1}{4} \: \Var_\gc(N)
      &
      \text{ grand canonical,}
    \end{array}
  \right.
\end{equation}
where
\begin{equation}
  \Ft \left(\y = \frac{T_F}{T} \right)
  = \frac{1-\e^{-y}}{4y}
  \:.
\end{equation}
In the thermal regime, we have not included the contribution of the quantum fluctuations at zero temperature, as it is subdominant, of order $\ln N$. 
We checked that these two regimes smoothly
match together for $T_Q \ll T \ll T_F$. Plots of these different
expressions for the variance are shown in Fig.~\ref{fig:PltVar}.

We stress an important difference between the two scaling functions
$\Fq$ and $\Ft$: while the latter is not universal, i.e. depends on
the precise form of the confining potential, the function $\Fq$ is
universal.

\subsection{Outline of the paper}

The paper is organised as follows: Section~\ref{sec:LinStat}
introduces the linear statistics for fermions. A general formula for
the variance of linear statistics is obtained. In
Section~\ref{sec:OccNb} we derive universal relations on the
occupation numbers, which are useful to derive our results for the
variance. We check our formulae in Section~\ref{sec:CheckKinEner} by
recovering the results of Ref.\cite{GreMajSch17} on the potential
energy. The case of the index variance for fermions in a harmonic well
is discussed in Section~\ref{sec:IndexFerm}. We end the paper with
some concluding remarks.  Some technical details are relegated to
Appendices.

\section{Observables of the form of linear statistics}
\label{sec:LinStat}

Consider a system of spinless and non-interacting fermions in a
confining potential $V(x)$. We denote $\psi_k(x)$ the one-particle
wave-function associated to the energy $\varepsilon_k$, with $k \in
\mathbb{N}$. These energy levels are non-degenerate in one
dimension. In the absence of interaction, a many-body quantum state
can be conveniently labelled by the set of occupation numbers $\{ n_k
\}$, where $n_k = 0$ or $1$ for fermions.  The total number of
fermions is then $ N = \sum_{k} n_k$ fermions. The associated
many-body wave function takes the form of a Slater determinant
combining the one-particle wave functions of the occupied levels:
\begin{equation}
  \Psi_{ \{ n_k \}  }(x_1, \ldots, x_N)
  = \frac{1}{\sqrt{N!}}
  \det[\psi_{k_i}(x_j)]_{1 \leq i,j \leq N}
  \:,
\end{equation}
where $\{ k_i \}_{i=1,\ldots,N}$ is the list of occupied levels:
$n_k = 1$ if $k \in \{ k_i \}$.

We will consider the situation where the fermions are subjected to
thermal fluctuations. In the following, we describe these fluctuations
either in the \textit{canonical} or the \textit{grand canonical}
ensemble.
Since the description of our system of fermions involves two different
sets of random variables, the positions and the occupation numbers, we
are led to define two types of averaging:
\begin{itemize}
\item the ``quantum average'', denoted $\moyQ{\cdots}_{\ket{ \{ n_k \}
    }}$, which corresponds to averaging over the positions of the
  fermions. This averaging is defined for a given quantum state
  $\ket{\{ n_k \}}$, with fixed number $N = \sum_k n_k$ of particles,
  as:
  \begin{equation}
    \label{eq:defQuantAverg}
    \moyQ{ F(x_1, \ldots, x_p) }_{\ket{ \{ n_k \}  }}
    = \int \dd x_1 \ldots \dd x_N \:
    \abs{\Psi_{\{ n_i \}}(x_1, \ldots, x_N)}^2
    \: F(x_1, \ldots, x_p)
    \:,
  \end{equation}
  for any function $F$ of $p \leq N$ positions. In the following we
  will often omit the subscript $\ket{\{ n_k \}} $ for simplicity.
\item the ``thermal average'', which
  corresponds to averaging over the occupation numbers. This averaging
  depends on the ensemble under consideration. We denote it
  $\moyTc{\cdots}$ in the canonical ensemble and $\moyTg{\cdots}$ in
  the grand canonical ensemble. For any function $G$ of the occupation
  numbers, it is defined as
  \begin{equation}
    \label{eq:defThermAverg}
    \moyT{G(\{ n_k \})} = 
    \sum_{\{ n_k \}} \mathscr{P}_{\rm c,g}(\{ n_k \}) \:
    G(\{ n_k \})
    \:,
  \end{equation}
  where $\mathscr{P}_{\rm c,g}(\{ n_k \})$ represents the canonical or
  grand canonical measures given by Eqs.~(\ref{eq:MeasureCan}) and
  (\ref{eq:MeasureGC}) respectively.
\end{itemize}
The quantum and thermal average will involve first averaging over the
positions of the fermions $\{ x_n \}$, Eq.~\eqref{eq:defQuantAverg}, and then over the occupation numbers, Eq.~\eqref{eq:defThermAverg}.
At zero temperature, the system is frozen in its ground state $\Psi_0$
and only the quantum average remains:
\begin{equation}
  \left. \moyT{\moyQ{ F(x_1, \ldots, x_p) }} \right|_{T=0}
  =  \int \dd x_1 \ldots \dd x_N \:
  \abs{\Psi_{0}(x_1, \ldots, x_N)}^2
  \: F(x_1, \ldots, x_p)
  \:.
\end{equation}
Such integrals can be evaluated by making use of the determinantal
structure of $\Psi_0$, as we will see in
section~\ref{sec:QuantAvrAndDet}. When going to finite temperature,
the excited states contribute to the thermal averaging:
\begin{align}
  \moyT{\moyQ{ F(x_1, \ldots, x_p) }}
  = \sum_{\{ n_k \}} 
  \mathscr{P}_{\rm c,g}(\{ n_k \})
  \int \dd x_1 \ldots \dd x_N \:
  \abs{\Psi_{\{ n_i \}}(x_1, \ldots, x_N)}^2
  \: F(x_1, \ldots, x_p)
  \:.
\end{align}
The integral over the positions can still be computed by using the
determinantal structure, but the summation over the quantum states
makes the problem much more challenging.

In the following, we will study a wide class of observables which take
the form of linear statistics $\L$ of positions of the
fermions, Eq.~(\ref{eq:defLinStat}).  Our aim is to study the variance
\begin{equation}
  \Var_{\rm c,g}(\L) = \moyT{\moyQ{\L^2}}
  -  \left( \moyT{\moyQ{\L}} \right)^2
  \:.
\end{equation}
In order to compute this variance, we first
need to evaluate the quantum averages $\moyQ{\L} $ and $\moyQ{\L^2}$ in
a given quantum state $\ket{\{ n_k \}}$. This is the object of the next
section. The thermal averaging will be discussed in Section~\ref{sec:OccNb}.

\subsection{Quantum averages and determinantal structure}
\label{sec:QuantAvrAndDet}

Let us first consider a quantum state $\ket{\{ n_k \}}$, which
contains $N = \sum_k n_k$ particles. It is convenient to use the fact
that the positions of the fermions, in that given state, is a
determinantal point process~\cite{DeLDoMajSch15,DeLDoMajSch16}. This
means that the modulus square of the many-body wave function can be
rewritten as a determinant
\begin{equation}
  \abs{\Psi_{\{n_k\}}(x_1,\ldots,x_N)}^2 =
  \frac{1}{N!}
  \det\left[ K(x_i,x_j;\{n_k\}) \right]_{1 \leq i,j \leq N}
  \:,
\end{equation}
where we have introduced the kernel
\begin{equation}
  \label{eq:Kernel}
   K(x,y;\{n_k\}) =
   \sum_{k=0}^{\infty} n_k \: \psi_{k}^*(x) \psi_k(y)
   \:.
\end{equation}
Since the wave functions are orthogonal, this kernel verifies the reproducibility property
\begin{equation}
  \int K(x,y;\{n_k\}) K(y,z;\{n_k\}) \dd y
  = K(x,z;\{n_k\}) \:,
\end{equation}
where we used that $n_k^2 = n_k$ since $n_k = 0$ or $1$ for fermions.
The direct consequence of this property is that the $n$-points
correlation function also takes a simple determinantal form:
\begin{align}
  \nonumber
  R_n(x_1,\ldots,x_n) 
  &= \frac{N!}{(N-n)!} \int \dd x_{n+1} \ldots \dd x_N
  \abs{\Psi_{\{n_k\}}(x_1,\ldots,x_N)}^2
  \\[0.2cm]
  &=  \det\left[ K(x_i,x_j;\{n_k\}) \right]_{1 \leq i,j \leq n} \:.
  \label{eq:nPtCorrFct}
\end{align}
In particular, the one-point function is given by
\begin{equation}
  \label{eq:OnePtCorrFct}
  R_1(x)
  = N \int \dd x_2 \ldots \dd x_N \:
   \abs{\Psi_{\{n_k\}}(x_1,\ldots,x_N)}^2
  = K(x,x;\{ n_k \})
  \:.
\end{equation}
When averaged over $n_k$'s in the grand canonical ensemble, using
Eq.~\eqref{eq:MeasureGC}, this is precisely the mean density of
fermions
\begin{equation}
\moyTg{\moyQ{\rho(x)}}=K(x,x;\{ \overline{n_k}^\gc \})
\end{equation}
where $\rho(x)=\sum_n\delta(x-x_n)$ and $\moyTg{\moyQ{\cdots}}$ is the
usual quantum and statistical average. The two-point correlation
function reads:
\begin{equation}
  \label{eq:TwoPtCorrFct}
  R_2(x,y) =  K(x,x;\{ n_k \}) K(y,y;\{ n_k \})
  - K(x,y;\{ n_k \})^2
  \:.
\end{equation}
When averaged over $n_k$'s in the grand canonical ensemble, this is
related to the familiar relation for the density-density correlation
function in the Fermi gas
\begin{equation}
  \label{eq:DensityDensityCorrelations}
  \moyTg{\moyQ{\rho(x)\,\rho(y)}}_\mathrm{corr}
  =\delta(x-y)\, \moyTg{\moyQ{\rho(x)}}
  - \Big|
    K(x,y;\{ \overline{n_k}^\gc \})
  \Big|^2
  \:,
\end{equation}
where the minus sign is related to the effective repulsion between
fermions due to the Pauli principle.  When the density can be
considered constant, equal to $\bar\rho$, the zero temperature kernel
is the famous sine-kernel
$K(x,y)|_{T=0}=\bar\rho\:\mathrm{sinc}[\bar\rho\pi(x-y)]$, where
$\mathrm{sinc}(z)=\sin z/z$, related to the anti-correlations
\begin{equation}
  \label{eq:BulkCorrelFct}
  \moyQ{\rho(x)\,\rho(y)}_\mathrm{corr}
  =\bar\rho\,\delta(x-y)-\bar\rho^2\,\mathrm{sinc}^2[\bar\rho\pi(x-y)]
  \qquad
  \text{(in bulk).}
\end{equation}

Using these properties we can easily express the quantum average of
the linear statistics~(\ref{eq:defLinStat}):
\begin{equation}
  \moyQ{\L}\Qs
  = \moyQ{\sum_{n=1}^N \ls(x_n) }
  = N \moyQ{\ls(x_1)}
  \:,
\end{equation}
where we used the symmetry under the exchange of particles.
Therefore, only the one-point function, Eq.~(\ref{eq:OnePtCorrFct}),
is needed for this computation:
\begin{equation}
  \moyQ{\L}\Qs
  = \int \ls(x) R_1(x) \dd x
  = \int \ls(x) K(x,x,\{n_k\}) \dd x
  \:.
\end{equation}
Using the expression of the kernel, Eq.~(\ref{eq:Kernel}), one can
express this average in terms of the one-particle wave functions:
\begin{equation}
  \label{eq:QuAvL0}
  \moyQ{\L}\Qs = \sum_k n_k \int \ls(x) \abs{\psi_k(x)}^2 \dd x \:.
\end{equation}
We can similarly compute the mean square:
\begin{align}
  \nonumber
  \moyQ{\L^2}\Qs
  &= N \moyQ{\ls(x_1)^2} + N(N-1) \moyQ{\ls(x_1)\ls(x_2)}
  \\
  &= \int \ls(x)^2 R_1(x) \: \dd x
  + \int \ls(x) \ls(y) R_2(x,y) \: \dd x \dd y
  \nonumber
  \\
  &= \sum_k n_k B_k
  + \sum_{k,l} n_k n_l (A_{k,k}A_{l,l} - (A_{k,l})^2 )
  \label{eq:QuAvL2}
  \:,
\end{align}
where we have introduced the matrix elements
\begin{equation}
  \label{eq:DefAklBk}
    A_{k,l} = \int \ls(x) \psi_k^*(x) \psi_l(x) \dd x \:,
    \quad
    B_k = \int \ls(x)^2 \abs{\psi_k(x)}^2 \dd x \:.
\end{equation}
Using these definitions, we can rewrite (\ref{eq:QuAvL0}) as:
\begin{equation}
  \label{eq:QuAvL}
  \moyQ{\L}\Qs = \sum_k n_k A_{k,k}
  \:.
\end{equation}
This derivation shows that one only needs to compute the matrix
elements $A_{k,l}$ and $B_k$ to perform the quantum averages involved
in the variance of a linear statistics $\mathcal{L}$.

\subsection{A general formula for the variance}

Using the results of the quantum average,
Eqs.~(\ref{eq:QuAvL2},\ref{eq:QuAvL}), one can straightforwardly take
the thermal average
\begin{equation}
  \moyT{\moyQ{\L}} = \sum_k \moyT{n_k} A_{k,k}
  \:,
\end{equation}
\begin{equation}
  \moyT{\moyQ{\L^2}} =
  \sum_k \moyT{n_k} B_k
  + \sum_{k,l} \moyT{n_k n_l} (A_{k,k}A_{l,l} - (A_{k,l})^2 )
  \:,
\end{equation}
in the canonical or grand canonical ensemble. Combining these two
relations, we obtain a general expression for the variance of a
linear statistics:
\begin{equation}
  \label{eq:VarLinStat}
  \boxed{
  \Var_{\rm c,g}(\L) = 
  \sum_k \moyT{n_k} B_k
  - \sum_{k, l} \moyT{n_k} \moyT{n_l} (A_{k,l})^2
  + \sum_{k\neq l} \Cov_{\rm c,g}(n_k,n_l) (A_{k,k} A_{l,l}- (A_{k,l})^2)
  }
\end{equation}
where $\Cov_{\rm c,g}(n_k,n_l) = \moyT{n_k n_l} - \moyT{n_k} \moyT{n_l}$. 
The difference between the two ensembles clearly appears on this
relation. Indeed, the covariance of occupation numbers is zero in the
grand canonical ensemble: $\Cov_\gc(n_k, n_l) = 0$ for $k \neq l$.  We
will see that this term gives an additional contribution of the same
order as the first two.

In order to use this formula for the variance, one first needs to
compute the coefficients $A_{k,l}$ and $B_k$, which will be analysed
in Appendix~\ref{App:QuantAvrIndex}. Then, the sums over the levels
need to be evaluated. An important piece of the calculation is a relation
on the occupation numbers which will be derived in the next section.

\section{Occupation numbers}
\label{sec:OccNb}

In this section, we derive general relations involving the occupation
numbers, valid both in the canonical and grand canonical ensembles.
Although this article focuses on fermionic systems, we will also
consider the bosonic case for the sake of generality.

\subsection{Grand canonical ensemble}

Let us first consider the case of the grand canonical ensemble. In
this ensemble, the mean occupation numbers are the well-known
Bose-Einstein and Fermi-Dirac distributions:
\begin{equation}
  \label{eq:MeanOccupNbGC}
  \moyTg{n_k} = \left\lbrace
    \begin{array}{ll}
      \displaystyle
      \frac{1}{\e^{\beta(\varepsilon_k - \mu)} - 1} & \text{ for
        bosons, with }
      \mu < \varepsilon_0
      \\[0.2cm]
      \displaystyle
      \frac{1}{\e^{\beta(\varepsilon_k - \mu)} + 1} & \text{ for
        fermions}
    \end{array}
  \right.
\end{equation}
where $\mu$ is the chemical potential, which controls the mean number
of particles
\begin{equation}
  \label{eq:linkMuNgc}
  \moyTg{N} = \sum_k \moyTg{n_k}
  \:.
\end{equation}
The variance of the occupation numbers is also known:
\begin{equation}
  \label{eq:VarNkGC}
  \Var_\gc(n_k)
  =  \moyTg{(n_k^2)} - (\moyTg{n_k})^2
  = \moyTg{n_k}(1 \pm \moyTg{n_k})
  \:,
\end{equation}
with the upper sign for bosons and the lower sign for fermions.
The simplicity and the success of the grand canonical ensemble relies
on the independence of individual energy levels, i.e. the absence of
correlations between occupation numbers:
\begin{equation}
  \Cov_\gc(n_k,n_l)
  =  \moyTg{n_k n_l} - \moyTg{n_k} \: \moyTg{n_l}
  = 0
  \:
  \quad \text{if} \quad
  k \neq l
  \:.
\end{equation}

\subsection{Canonical ensemble}

In the canonical ensemble, the total number of particles is fixed to
$N$. This constraint induces correlations between the occupation
numbers associated to different levels, which are in general very
difficult to handle. The mean canonical occupation numbers are
obtained by averaging with the canonical
measure~(\ref{eq:MeasureCan}):
\begin{equation}
  \label{eq:MeanOccCan0}
  \moyTc{n_k} = \sum_{\{ n_i \}}
  n_k \: \mathscr{P}_\can(\{ n_i \})
  = \frac{1}{Z_\can(N)}\sum_{\{ n_i \}}
  n_k
  \:
  \e^{-\beta \sum_i n_i \varepsilon_i}
  \:
  \delta_{\sum_i n_i, N}
  \:.
\end{equation}
The general strategy of statistical physics is to introduce generating
functions, i.e. consider the sum
\begin{equation}
  \label{eq:GenFctNk}
  \sum_{N=0}^\infty \fug^N 
  \sum_{\{ n_i \}}
  n_k
  \:
  \e^{-\beta \sum_i n_i \varepsilon_i}
  \:
  \delta_{\sum_i n_i, N}
  =
  Z_\gc(\fug) \sum_{\{ n_i \}}
  n_k \: \mathscr{P}_\gc(\{ n_i \})
  =  Z_\gc(\fug) \: \moyTg{n_k}
  \:,
\end{equation}
where we recognised the grand canonical measure,
Eq.~(\ref{eq:MeasureGC}). We can then deduce $\moyTc{n_k}$ by a
contour integral which selects the $\fug^N$ term in
Eq.~(\ref{eq:GenFctNk}):
\begin{equation}
  \moyTc{n_k} = \frac{1}{Z_\can(N)} 
  \oint \frac{\dd \fug}{2\I \pi} \frac{Z_\gc(\fug)}{\fug^{N+1}} 
  \moyTg{n_k}
  \:,
\end{equation}
where the integrals run over a closed contour winding once around the
origin in the counter-clockwise direction.
The partition functions $Z_\can$ and $Z_\gc$ being related by
\begin{equation}
  \label{eq:LinkZcZg}
  Z_\gc(\fug) = \sum_{N} \fug^N Z_\can(N)
  \:,
  \qquad
  Z_\can(N) = \oint \frac{\dd \fug}{2\I \pi}
  \frac{Z_\gc(\fug)}{\fug^{N+1}}
  \:,
\end{equation}
we can rewrite the occupation numbers as
\begin{equation}
  \label{eq:OintMeanNkCan}
  \moyTc{n_k} =
  \frac{
    \displaystyle
    \oint \frac{\dd \fug}{2\I \pi} \frac{Z_\gc(\fug)}{\fug^{N+1}} 
    \moyTg{n_k}
  }
  {
    \displaystyle
    \oint \frac{\dd \fug}{2\I \pi} \frac{Z_\gc(\fug)}{\fug^{N+1}}
  }
  \:.
\end{equation}
A similar relation clearly
holds for any average quantity, for instance
\begin{equation}
  \label{eq:OintCovNkCan}
  \moyTc{n_k n_l} = \frac{1}{Z_\can(N)} 
  \oint \frac{\dd \fug}{2\I \pi} \frac{Z_\gc(\fug)}{\fug^{N+1}} 
  \moyTg{n_k n_l}
  = \frac{
    \displaystyle
    \oint \frac{\dd \fug}{2\I \pi} \frac{Z_\gc(\fug)}{\fug^{N+1}} 
    \moyTg{n_k n_l}
  }
  {
    \displaystyle
    \oint \frac{\dd \fug}{2\I \pi} \frac{Z_\gc(\fug)}{\fug^{N+1}}
  }
  \:.
\end{equation}
We will first derive general expressions
for~(\ref{eq:OintMeanNkCan},\ref{eq:OintCovNkCan}). In a second step
we will analyse the large $N$ limit by a saddle point method. The two
results will be useful in the next sections.

\subsubsection{General representation for the covariance}

To evaluate the numerator in Eq.~(\ref{eq:OintMeanNkCan}), we only
need to determine the coefficient of the term $\fug^N$ in the
expansion of $Z_\gc(\fug) \moyTg{n_k}$ as a power series in
$\fug$. This series can be obtained by using
Eqs.~(\ref{eq:MeanOccupNbGC},\ref{eq:LinkZcZg}). Then, isolating the
term proportional to $\fug^N$ yields:
\begin{equation}
  \label{eq:OccupCan}
    \moyTc{n_k} = \sum_{p=1}^N (\pm 1)^{p-1} \frac{Z_\can(N-p)}
    {Z_\can(N)}
    \:
    \e^{-p \beta \varepsilon_k}
    \:,
\end{equation}
where the upper sign is for bosons, and the lower sign for
fermions. This relation was known in the literature, see
e.g. Refs.\cite{WeiWil97,BorHarMulHil99,ChaMekZam99}. Similarly, we
can obtain the expression for the product of occupation numbers:
\begin{equation}
  \label{eq:CovCan}
    \moyTc{n_k n_l} = 
    \sum_{p=1}^{N-1} \sum_{q=1}^{N-p-1} (\pm 1)^{p+q} 
    \frac{Z_\can(N-p-q)}{Z_\can(N)}
    \:
    \e^{-p\beta \varepsilon_k} \: \e^{-q \beta \varepsilon_l}
    \:,
    \quad
    \text{for}
    \quad
    k \neq l
    \:.
\end{equation}
This last relation can be found in the case of bosons in
Ref.~\cite{ChaMekZam99}. It can be simplified by introducing $s=p+q$:
\begin{align}
  \nonumber
  \moyTc{n_{k}n_{l}}
  &= \sum_{s=2}^{N-1} (\pm 1)^s 
  \frac{Z_\can(N-s)}{Z_\can(N)} \: \e^{-\beta s \varepsilon_l}
  \sum_{p=1}^{s-1} \e^{-\beta p (\varepsilon_k - \varepsilon_l)}
  \\
  &= \sum_{s=2}^{N} (\pm 1)^s \frac{Z_\can(N-s)}{Z_\can(N)}
  \:
  \frac{\e^{-\beta (s-1) \varepsilon_l}-\e^{-\beta (s-1)\varepsilon_{k}}}
  {\e^{\beta \varepsilon_k}-\e^{\beta \varepsilon_l}}
  \:.
\end{align}
Separating this expression into two sums, we recognise the expressions
of $\moyTc{n_k}$ and $\moyTc{n_l}$, Eq.~(\ref{eq:OccupCan}), thus,
\begin{equation}
  \label{eq:RelCovOccNbCan}
    \moyTc{n_{k}n_{l}}
    = (\mp) \frac{
      \e^{\beta \varepsilon_k} \: \moyTc{n_{k}}
      -\e^{\beta \varepsilon_l} \: \moyTc{n_{l}}
    }{\e^{\beta \varepsilon_k}-\e^{\beta \varepsilon_l}}
    \:.
\end{equation}
We derived this relation in the canonical ensemble, but one can easily
check that it also holds in the grand-canonical one, where it becomes
$\moyTg{n_k n_l} = \moyTg{n_k} \moyTg{n_l}$.  Therefore, this is a
very general relation, valid for any system of non interacting bosons
or fermions, either in the canonical or grand canonical ensemble. To
the best of our knowledge, this relation \eqref{eq:RelCovOccNbCan}
involving the occupation numbers is new.

We have extended this analysis to the $p$-point correlation functions
in Ref.~\cite{GirGraTex17}.

\subsubsection{Saddle point estimate for large $N$}
\label{subsec:SaddlePt}

Despite having obtained general
relations~(\ref{eq:OccupCan},\ref{eq:CovCan}), their large $N$
analysis remains a challenge. For this purpose, we perform a saddle
point analysis of Eqs.~(\ref{eq:OintMeanNkCan},\ref{eq:OintCovNkCan}).
For all the integrals, the saddle point $\fug_\star$ is given by
the condition
\begin{equation}
  \sum_k \moyTg{n_k} = N
  \:,
\end{equation}
which fixes the fugacity (or the chemical potential) such that the
grand canonical mean number of particles in the trap is equal to the
fixed canonical number $N$. Then, the integral
(\ref{eq:OintMeanNkCan}) yields:
\begin{equation}
  \moyTc{n_k} = \moyTg{n_k} + \O(N^{-1})
  \:.
\end{equation}
This indicates that the statistics of the occupation numbers are the
same at leading order in $N$ in the canonical and grand-canonical
ensembles. This is not surprising since we expect both ensembles to be
equivalent for averaged quantities in the thermodynamic limit.  The
equivalence of thermodynamic results however only holds for averages
and not for variances or covariances, as it is
well-known~\cite{PatBea11}. Performing the same saddle point analysis
with Eq.~(\ref{eq:OintCovNkCan}) yields
\begin{equation}
  \moyTc{n_k n_l} = \moyTg{n_k} \: \moyTg{n_l}  + \O(N^{-1})
  \:.
\end{equation}
Since we will need the covariance of occupation numbers, we push the
saddle point approximation of
Eqs.~(\ref{eq:OintMeanNkCan},\ref{eq:OintCovNkCan}) to the next order,
see Appendix~\ref{app:SaddlePt}, in order to get the $\O(N^{-1})$
corrections. After many simplifications, we obtain the compact
expression:
\begin{equation}
  \label{eq:CovOccNbCanSadPt}
  \Cov_\can(n_k,n_l)
  = - \frac{\Var_\gc(n_k) \Var_\gc(n_l)}
  {\sum_p \Var_\gc(n_p)}
  + \O(N^{-2})
  \:,
\end{equation}
where $\Var_\gc(n_k)$ is given by Eq.~(\ref{eq:VarNkGC}). Note that
since $\Var_\gc(n_k) = \Var_\can(n_k) + \O(N^{-1})$, we can express this
covariance in terms of the variance of occupation numbers in any
ensemble. We chose to express it in terms of the grand canonical one
since the expressions are simpler in this case. 
In the  canonical ensemble, $N=\sum_kn_k$ is fixed, which implies the sum rule
\begin{equation}
  \sum_k \Var_\can(n_k) + \sum_{k \neq l}
  \mathrm{Cov}_\can(n_k,n_l)
  = 0
  \:.
\end{equation}
It is clear that Eq.~\eqref{eq:CovOccNbCanSadPt} is consistent with
this sum rule, up to~$\mathcal{O}(N^0)$.

However relation~(\ref{eq:CovOccNbCanSadPt}) is valid as long as the
covariance is a $\O(N^{-1})$ correction to the saddle point
approximation. It is the case when the denominator, which corresponds
to the variance of the total number of particles in the
grand-canonical ensemble, is of order $N$: $\sum_p \Var_\gc(n_p) =
\Var_\gc(N) = \O(N)$. As discussed in Appendix~\ref{app:VarN}, this is
verified only in the thermal regime $T \sim T_F$. Therefore, this
relation for the covariance can only be used this regime.  In the
quantum regime $T \sim T_Q$ we will instead rely on the exact
relation~(\ref{eq:RelCovOccNbCan}).


\subsection{A symmetry relation for fermion occupation numbers}

Let us now consider fermions in a harmonic trap $V(x)= \frac{1}{2} m
\omega^2 x^2$, which will be the main focus of the paper. In this
case, the spectrum is linear, $\varepsilon_n = (n+\frac{1}{2})\hbar
\omega$, $n \in \mathbb{N}$.

In the grand canonical ensemble, the Fermi-Dirac
distribution~(\ref{eq:MeanOccupNbGC}) has a the well-known symmetry
around the chemical potential $\mu$:
\begin{equation}
  \frac{1}{\e^{\beta(\varepsilon-\mu)}+1}
  = 1 - \frac{1}{\e^{-\beta(\varepsilon-\mu)}+1}
  \:,
\end{equation}
which is usually interpreted as the particle-hole symmetry.
In the case of a discrete spectrum, setting the chemical potential in the middle of a gap, $\mu = N_f \hbar \omega$, where $N_f$ is an integer, the relation reduces to:
\begin{equation}
  \label{eq:SymOcNbFermGC}
  \moyTg{n_{N_f+k}} = 1 - \moyTg{n_{N_f-k-1}}
  \:.
\end{equation}
Due to the linearity of the spectrum, we have $\moyTg{N} = N_f$ up to
exponentially small correction $\sim(\kB
T/\hbar\omega)\,\e^{-N_f\hbar\omega/(\kB T)}$.

In the canonical ensemble the occupation numbers~(\ref{eq:OccupCan}),
are expressed in terms of the canonical partition function. For the
harmonic oscillator, it can be computed analytically, and the result
can be found in a few textbooks~\cite{TodKubSai83,TexRou17}:
\begin{equation}
  \label{eq:ZcanoOscillators}
  Z_\can(N) = \e^{-\frac{N^2 \beta \hbar \omega}{2}}
  \prod_{n=1}^N \frac{1}{1-\e^{-n\beta \hbar \omega}}
  \:.
\end{equation}
Using this result, the expression of the occupation numbers
(\ref{eq:OccupCan}) greatly simplifies near the Fermi level in the
large $N$ limit:
\begin{equation}
  \moyTc{n_{N+k}} \simeq 
  \sum_{p=1}^\infty (-1)^{p-1}
  \:
  \e^{-\frac{\beta \hbar \omega}{2} \: p(2k+p+1)}
  \:,
\end{equation}
with corrections exponentially small with $N$. Using this expression,
it is straightforward to show that the canonical occupation numbers
also exhibit the particle-hole symmetry around the Fermi level:
\begin{equation}
  \label{eq:SymOcNbFerm}
  \moyTc{n_{N+k}} = 1 -  \moyTc{n_{N-k-1}}
  \:,
\end{equation}
which is exactly the same as the grand canonical
relation~(\ref{eq:SymOcNbFermGC}).

These symmetries~(\ref{eq:SymOcNbFermGC},\ref{eq:SymOcNbFerm}), along
with relations~(\ref{eq:RelCovOccNbCan},\ref{eq:CovOccNbCanSadPt}) on
the occupation numbers will be essential for our study of linear
statistics for fermions in a harmonic trap. We will first check our
approach by recovering the recent results of Ref.~\cite{GreMajSch17}
on the potential (or kinetic) energy of the fermions. Then, we will
apply our method to the observable introduced in
Section~\ref{sec:Introduction}: the number $\N$ of fermions on the
positive axis, given by Eq.~(\ref{eq:DefNp}).


\section{A first check: potential energy of fermions in a harmonic trap}
\label{sec:CheckKinEner}

In the harmonic trap, the one-particle wave functions are expressed in
terms of Hermite polynomials:
\begin{equation}
  \label{eq:WaveFctHarmOsc}
  \psi_n(x) = \sqrt{\frac{\alpha}{2^n n! \sqrt{\pi}} } \:
  H_n(\alpha x) \: \e^{-\alpha^2 x^2/2} \:,
  \quad
  \alpha = \sqrt{\frac{m \omega}{\hbar}} \:,
\end{equation}
with energies $\varepsilon_n = (n+1/2)\hbar \omega$, $n \in
\mathbb{N}$. We denote $\{ x_n \}$ the positions of the fermions.

Recently, the distribution of the potential energy $E_p$, or
equivalently the kinetic energy, was obtained in
Ref.~\cite{GreMajSch17} for any temperature $T$ or fixed number $N$ of
fermions (canonical ensemble).
However their method is restricted to the study of the potential
energy
\begin{equation}
  E_p = \frac{1}{2} m \omega^2 I
  \:,
  \qquad
  I = \sum_{n=1}^N x_n^2
  \:,
\end{equation}
which is a specific linear statistics~(\ref{eq:defLinStat}), with
$\ls(x) = x^2$.  They have obtained two different scaling functions
describing the quantum and thermal regimes:
\begin{align}
  \label{eq:VarKinEQuant}
  \Var_\can(I) &\simeq \frac{N^2}{2 \alpha^4}\,
  V_\mathrm{q}\left(\frac{T}{T_Q}\right) \hspace{0.5cm}\mbox{for } T
  \sim T_Q
\end{align}
and 
\begin{align}
  \label{eq:VarKinETherm}
  \Var_\can(I) \simeq \frac{N^3}{2 \alpha^4 }\,
  V_\mathrm{th}\left(\frac{T}{T_F}\right) \hspace{0.5cm}\mbox{for } T
  \sim T_F
\end{align}
where the two scaling functions are 
\begin{align}
  \label{eq:VqJacek}
  V_\mathrm{q}\left( z \right)
  &= \coth\frac{1}{z}
  \:,
  \\
  \label{eq:VthJacek}
  V_\mathrm{th}\left( z \right)
  &=z \, \left[
    -6z^2 \, \mathrm{Li}_2(1-\e^{1/z})
    -1 - \coth\frac{1}{2z}
  \right]
  \:,
\end{align}
where $\mathrm{Li}_s(z) = \sum_{k=0}^\infty z^k/k^s$ is the
polylogarithm function. One can check that these two expressions
(\ref{eq:VarKinEQuant},\ref{eq:VarKinETherm}) smoothly match in the
intermediate regime $T_Q\ll T\ll T_F=NT_Q$, as the two scaling
function present the limiting behaviours $V_\mathrm{q}\left( z
\right)\simeq z$ for $z\to\infty$ and $V_\mathrm{th}\left( z
\right)\simeq z$ for $z\to0$.  In this section we will check using our
more general approach that we recover these results.

In this particular case of linear statistics with $\ls(x) = x^2$, the
matrix elements~(\ref{eq:DefAklBk}) can be computed exactly:
\begin{equation}
  \label{eq:BkKinE}
  B_k = \frac{3}{2 \alpha^4}
  \left( k^2 + k+ \frac{1}{2} \right)
  \:,
\end{equation}
\begin{equation}
  \label{eq:AkKinE}
  A_{k,l} = \frac{1}{\alpha^2}
  \left\lbrace
    \left( k + \frac{1}{2} \right)  \delta_{k,l}
    + \frac{1}{2} \sqrt{(k+1)(k+2)} \: \delta_{k+2,l}
    + \frac{1}{2} \sqrt{k(k-1)} \: \delta_{k-2,l}
    \right\rbrace
    \:.
\end{equation}

\subsection{Quantum regime}
\label{sec:QuantRegKinE}

In this regime, the small temperature $T \sim T_Q$ allows only a few
excitations above the Fermi level $\varepsilon_N$. Therefore, we
expect the main contribution to come from the proximity of this
level. At leading order in $N$, for fixed $k$ and $l$, the matrix
elements become:
\begin{equation}
  B_{N+k} \simeq \frac{3N^2}{2 \alpha^4}
  \:,
\end{equation}
\begin{equation}
  A_{N+k,N+l} \simeq \frac{N \hbar}{m \omega}
  \left\lbrace
    \delta_{k,l}
    + \frac{1}{2} \: \delta_{k+2,l}
    + \frac{1}{2} \: \delta_{k-2,l}
    \right\rbrace
    \:.
\end{equation}
Using these expressions in Eq.~(\ref{eq:VarLinStat}), the variance of
$I$ becomes, at leading order:
\begin{equation}
  \Var_\can(I) \simeq \frac{N^2}{4 \alpha^4}
  \sum_{k=-\infty}^\infty
  \left(
    2 \: \moyTc{n_{N+k}} - \moyTc{n_{N+k}n_{N+k-2}}
    - \moyTc{n_{N+k}n_{N+k+2}}
  \right)
  \:,
\end{equation}
where we extended the summation to $-\infty$ instead of $-N$, since
the corrections are exponentially small. Using then
relation~(\ref{eq:RelCovOccNbCan}), this becomes
\begin{equation}
  \Var_\can(I) \simeq \frac{N^2}{4 \alpha^4}
  \coth(\beta \hbar \omega)
  \sum_{k=-\infty}^{+\infty}
  (\moyTc{n_{N+k}}- \moyTc{n_{N+k+2}})
  \:.
\end{equation}
This last sum cannot be separated into two sums because they would
both diverge. Using the symmetry of the mean occupation numbers around
the Fermi level, see Eq.~(\ref{eq:SymOcNbFerm}), we get
\begin{equation}
  \sum_{k=-\infty}^{\infty} (\moyTc{n_{N+k}}-\moyTc{n_{N+k+2}})
  =
  \moyTc{n_{N-2}} - \moyTc{n_{N}} + \moyTc{n_{N-1}}- \moyTc{n_{N+1}}
  + 2 \sum_{k=0}^\infty (\moyTc{n_{N+k}}-\moyTc{n_{N+k+2}})
  \:.
\end{equation}
Under this form, the sum can be separated into two sums. This yields
\begin{equation}
  \sum_{k=-\infty}^{\infty} (\moyTc{n_{N+k}}-\moyTc{n_{N+k+2}})
  = 2 ( \moyTc{n_{N}} +  \moyTc{n_{N+1}}) + \moyTc{n_{N-2}} - \moyTc{n_{N}} + \moyTc{n_{N-1}}- \moyTc{n_{N+1}}
  \:.
\end{equation}
Finally, using again Eq.~(\ref{eq:SymOcNbFerm}) gives
\begin{equation}
  \sum_{k=-\infty}^{\infty} (\moyTc{n_{N+k}}-\moyTc{n_{N+k+2}})
  = 2
  \:.
\end{equation}
Therefore, we recover the result of Ref.~\cite{GreMajSch17} in the
quantum regime, Eq.~(\ref{eq:VarKinEQuant}):
\begin{equation}
  \Var_\can(I) \simeq \frac{N^2}{2 \alpha^4}
  \coth(\beta \hbar \omega)
  \:.
\end{equation}

\subsection{Thermal regime}
\label{sec:ThermRegKinE}

We now consider the regime where the temperature is of the order of
the Fermi temperature $T \sim T_F$. We fix $\y = \beta N \hbar
\omega = T_F/T$ and let $N \to \infty$. In this case, we can use the results
of section~\ref{subsec:SaddlePt}, for instance
\begin{equation}
  \label{eq:MoyNkGCbis}
  \moyTc{n_k} \simeq \moyTg{n_k} + \O(N^{-1})
  = \frac{1}{\e^{\beta(\varepsilon_k - \mu)} + 1} + \O(N^{-1})
  \:,
\end{equation}
where the chemical potential $\mu$ is fixed by $ \sum_k \moyTg{n_k} =
N $. This sum over $k$ can be replaced by an integral over $x=k/N$ since
the discreteness of the spectrum plays no role in this regime:
\begin{equation}
  \sum_k  \moyTg{n_k} \simeq
  N \int_0^\infty \frac{\dd x}{\e^{\y (x - \mu/(N \hbar \omega)  )}+1}
  = \frac{N}{\y} \ln \left(
    1 + \e^{\y \mu/N \hbar \omega}
  \right)
  \:.
\end{equation}
Imposing that this sum is the total number of particles
yields:~\footnote{ A more precise treatement of the steepest descent
  equation $\sum_k \moyTg{n_k} = N$, with the help of the
  Euler-MacLaurin formula $\sum_{k=0}^\infty f(k)=\int_0^\infty\dd
  x\,f(x) - \sum_{p=1}^\infty\frac{B_p}{p!}\,f^{(p-1)}(0)$ with $B_p$
  the Bernoulli number ($B_1=-1/2$, $B_2=1/6$, etc), shows that the
  zero temperature limit of the chemical potential is precisely in the
  middle of the energy gap above the last occupied level: $\mu = N
  \hbar
  \omega+\mathcal{O}(N^{-1})=(\varepsilon_{N}+\varepsilon_{N-1})/2$.
}
\begin{equation}
  \label{eq:ChemPotThermReg}
  \mu =  N  \hbar \omega
  + \frac{N \hbar \omega}{\y} \ln(1 - \e^{-\y})
  + \O(N^0)
  \:,
  \quad
  \y = \frac{T_F}{T}
  \:.
\end{equation}
Therefore, we can rewrite the occupation numbers \eqref{eq:MoyNkGCbis}
as
\begin{equation}
  \label{eq:ApproxOccupNb}
  \moyTc{n_k} \simeq f_{\y}(k/N) + \O(N^{-1})
  \:,
\end{equation}
where we introduced the notation
\begin{equation}
  \label{eq:fyx}
  f_\y(x) = \frac{1}{\displaystyle
    \frac{\e^{\y(x-1)}}{1-\e^{-\y}} + 1
  }
\end{equation}
for the Fermi-Dirac distribution in terms of convenient variables.  In
most cases, the approximation~(\ref{eq:ApproxOccupNb}) is sufficient.
However, the corrections are essential when the fluctuations of the
occupation numbers contribute. For instance, the covariances in the
last term of Eq.~(\ref{eq:VarLinStat}) must be evaluated using
Eq.~(\ref{eq:CovOccNbCanSadPt}).  Replacing now the sums over $k$ by
integrals over $x=k/N$, and keeping only the leading order of the
matrix elements~(\ref{eq:BkKinE},\ref{eq:AkKinE}) yields:
\begin{align}
  \Var_\can(I) \simeq &
  \frac{N^3}{\alpha^4}
  \int_0^\infty \left(
    \frac{3}{2} f_\y (x)
    - f_\y (x)^2
  \right) x^2 \: \dd x
  - \frac{N}{2\alpha^4} \int_0^\infty
    f_\y (x)^2 x^2 \: \dd x
  \nonumber
  \\
  & - \frac{N^3}{\alpha^4} \frac{
    \displaystyle
    \left(
      \int_0^\infty f_\y(x) (1-f_\y(x)) x \: \dd x
    \right)^2
  }{
    \displaystyle
    \int_0^\infty f_\y(x) (1-f_\y(x)) \: \dd x
  }
  \:.
\end{align}
Evaluating these integrals, we finally obtain
\begin{equation}
  \Var_\can(I) \simeq \frac{N^3}{2 \alpha^4 \y}
  \left(
    - \frac{6}{\y^2} \mathrm{Li}_2(1-\e^{\y})
    - 1 - \coth \frac{\y}{2}
  \right)
  \hspace{0.5cm}\mbox{for } y = \frac{T_F}{T}  
  \:.
\end{equation}
Hence we recover the result of Ref.~\cite{GreMajSch17} in the thermal
regime, Eq.~(\ref{eq:VarKinETherm}). This verification validates our
approach to compute the variance of linear statistics.

\subsection{Discussion}

It is interesting to comment on the physical content of these results,
and in particular compare the fluctuations of the potential energy
with the fluctuations of the total energy $E=E_c+E_p$, which, in the
canonical ensemble, can be related to the heat capacity studied in
detail in \cite{TexRou17} by $\mathrm{Var}_\can(E) = \kB T^2\,C_V$.
From \eqref{eq:ZcanoOscillators}, we get
\begin{equation}
  \label{eq:VarTotalEnergyCanonical}
   \mathrm{Var}_\can(E)
   = \sum_{n=1}^N  \left(\frac{n\hbar\omega/2}{\sinh(n\beta\hbar\omega/2)}\right)^2
   \simeq 
    \begin{cases}
     (\hbar\omega)^2 \,\e^{-\hbar\omega/(\kB T)} & \mbox{for } T\ll T_Q
     \\
     \frac{\pi^2}{3}\frac{(\kB T)^3}{\hbar\omega}=\frac{\pi^2}{3} N (\kB T)^2\big(\frac{T}{T_F}\big)
     & \mbox{for } T_Q\ll T\ll T_F 
     \\
     N (\kB T)^2 & \mbox{for }  T\gg T_F 
  \end{cases}
  \:.
\end{equation}
In the low temperature regime, the exponential suppression of the
fluctuations can be related to the existence of a gap $\hbar\omega$ in
the excitation spectrum.  The result in the intermediate regime
$T_Q\ll T\ll T_F$ has been rewritten in terms of the Fermi temperature
to make clear that it corresponds to the classical result
$\mathrm{Var}_\can(E)\simeq N (\kB T)^2$ (given by the equipartition
theorem) multiplied by the small factor $T/T_F\ll1$.  This well-known
suppression factor originates from the Pauli principle, which
restricts thermal fluctuations to take place in a small window of
width $\kB T$ around the Fermi level.

The fluctuations of the potential energy are given by
Eq.~\eqref{eq:VarKinEQuant} and \eqref{eq:VarKinETherm}:
\begin{equation}
   \mathrm{Var}_\can(E_p)    
   \simeq 
    \begin{cases}
     \frac18 N^2(\hbar\omega)^2 \,\left[1+2\e^{-2\hbar\omega/(\kB T)} \right]
     & \mbox{for } T\ll T_Q
     \\
     \frac18 N^2\hbar\omega\,\kB T  = \frac{1}{8}N (\kB T)^2\big(\frac{T_F}{T}\big)
       & \mbox{for } T_Q\ll T\ll T_F 
     \\
     \frac12 N (\kB T)^2 
     & \mbox{for }  T\gg T_F 
  \end{cases}
  \:.
\end{equation}
The finite value $\mathrm{Var}_\can(E_p)\sim(N\hbar\omega)^2$ at $T=0$
is a manifestation of the quantum fluctuations in the ground state,
like Eq.~\eqref{eq:VarT0}.  The relative classical fluctuations (for
$T\gg T_F$) behaves as
$\mathrm{Var}_\can(E_p)/\big(\overline{E_p}^\can\big)^2\simeq1/N$, as
expected, while the relative quantum fluctuations reach the value
$\mathrm{Var}_\can(E_p)/\big(\overline{E_p}^\can\big)^2=2/N^2$ at
$T=0$.  The comparison between the potential energy and the total
energy is more interesting: in the classical regime ($T\gg T_F$), we
have $\mathrm{Var}_\can(E_p) \simeq (1/2)\,\mathrm{Var}_\can(E)$, as
it should for the harmonic potential.  In the regime dominated by
quantum correlations ($T\ll T_F$), we have rather
$\mathrm{Var}_\can(E_p) \gg \mathrm{Var}_\can(E)$.  This observation
has interesting consequences for the correlations of kinetic and
potential energies.  We express the variance of the total energy
$E=E_c+E_p$, and use the fact that $E_c$ and $E_p$ have the same
statistical properties for harmonic confinement:
\begin{equation}
  \mathrm{Var}_\can(E) = 2\, \mathrm{Var}_\can(E_p)  + 2 \, \mathrm{Cov}_\can(E_c,E_p)
  \:.
\end{equation}
In the classical regime $T\gg T_F$, we have obtained
$\mathrm{Var}_\can(E_p)=\mathrm{Var}_\can(E_c)\simeq(1/2)\mathrm{Var}_\can(E)$,
which is related to the well-known fact that the kinetic and potential
energy are uncorrelated: $\mathrm{Cov}_\can(E_c,E_p)\simeq0$.  In the
regime $T\ll T_F$, we have obtained that
$\mathrm{Var}_\can(E_p)\gg\mathrm{Var}_\can(E)$, implying that
potential and kinetic energies are anti-correlated
\begin{equation}
  \frac{\mathrm{Cov}_\can(E_c,E_p)}{\sqrt{\mathrm{Var}_\can(E_c)\mathrm{Var}_\can(E_p)}} \simeq -1
  \hspace{1cm}\mbox{for }
  T\ll T_F
  \:,
\end{equation}
so that the fluctuations can be related as $\delta E_p\simeq -\delta
E_c$.

\section{Index variance for fermions in a harmonic trap}
\label{sec:IndexFerm}

We now apply the general considerations of sections~\ref{sec:LinStat}
and~\ref{sec:OccNb} to the study of the index $\N$, corresponding to
the number of fermions on the positive axis. It is given by
Eq.~(\ref{eq:DefNp}).  This quantity is a linear statistics: it is of
the form (\ref{eq:defLinStat}), with $\ls(x) = \Theta(x)$. Therefore,
we can use the results of section~\ref{sec:LinStat}. In particular,
the variance of $\N$ is given by Eq.~(\ref{eq:VarLinStat}), with
\begin{equation}
  \label{eq:AklHarmOsc}
  A_{k,l} = \int_{0}^{\infty} \psi_k(x) \psi_l(x) \: \dd x
  \:,
\end{equation}
and
\begin{equation}
  \label{eq:91}
  B_k = A_{k,k} = \int_{0}^{\infty} \psi_k(x)^2 \: \dd x
  = \frac{1}{2}
  \:.
\end{equation}
This last relation is exact, due to the symmetry of the potential.
Using this result, we can rewrite Eq.~(\ref{eq:VarLinStat}) as:
\begin{equation}
  \label{eq:92}
  \Var_{\rm c,g}(\N) = \frac{1}{4} \sum_k \moyT{n_k}
  + \frac{1}{4} \left( \sum_k \Var_{\rm c,g}(n_k) + 
    \sum_{k \neq l} \Cov_{\rm c,g} (n_k,n_l)  \right)
  - \sum_{k \neq l} \moyT{n_k n_l} (A_{k,l})^2
  \:.
\end{equation}
The first term gives the mean number of particles.  Moreover the
second term has a simple structure thanks to the fact that the matrix
elements $B_k$ and $A_{k,k}$ are equal and independent of $k$, in
Eq.~\eqref{eq:91}.  Note that this property in Eq.~\eqref{eq:91} is
specific to the choice of the observable considered here, namely the
index $\N$.  Writing $N=\sum_kn_k$, the second term can be simply
identified as the variance of the total number of particles
\begin{equation}
  \Var_{\rm c,g}(N) = \sum_k \Var_{\rm c,g}(n_k) + 
    \sum_{k \neq l} \Cov_{\rm c,g} (n_k,n_l) 
\end{equation}
where obviously $\Var_\can(N)=0$ by definition, while $\Var_\gc(N)$ is
finite.  Thus:
\begin{equation}
  \label{eq:VarIndexGen}
  \Var_{\rm c,g}(\N) = \frac{1}{4} \: \moyT{N}
  + \frac{1}{4} \: \Var_{\rm c,g}(N)
  - \sum_{k \neq l} \moyT{n_k n_l} (A_{k,l})^2
  \:.
\end{equation}
Before studying into detail the variance of $\N$ at any temperature,
we will first discuss the limit of high temperature in which the
fermions behave as classical particles.

\subsection{High temperature limit: the Maxwell-Boltzmann regime}
\label{subsec:MaxBoltz}

We start by considering the simplest limiting case, the limit of high
temperature $T \gg T_F$. In this case the fermions can be considered
as classical particles as the thermal fluctuations
dominate. Therefore, their positions $\{ x_n \}$ are independent, and
they follow the Maxwell-Boltzmann distribution:
\begin{equation}
  \mathrm{Proba}(x_n \in [x,x+\dd x]) 
  = \sqrt{\frac{\beta m \omega^2}{2\pi}} \: \e^{-\beta m\omega^2
    x^2/2} \: \dd x
  = p(x) \: \dd x
  \:.
\end{equation}
The probability that a particle is in the domain $x>0$ is $ p_{+} =
\frac{1}{2} $. We now need to distinguish the statistical ensembles:
\begin{itemize}
\item In the canonical ensemble, the number $N$ of particles in the
  trap is fixed. Therefore, the mean number of particles with position
  $x_n> 0$ is:
  \begin{equation}
    \label{eq:MeanNIMaxBoltzCan}
    \moyTc{\N} = N p_+ = \frac{N}{2}
    \:.
  \end{equation}
  To compute the variance, we also need the square of $\N$:
  \begin{equation}
    (\N)^2 = \left( \sum_n  \Theta(x_n) \right)^2
    = \sum_n  \Theta(x_n) + \sum_{n\neq m} \Theta(x_n) \Theta(x_m)
    \:,
  \end{equation}
  from which we deduce:
  \begin{equation}
    \label{eq:MeanNI2MaxBoltzCan}
    \moyTc{(\N)^2} = N p_+ + N(N-1) p_+^2
    = \frac{N(N+1)}{4}
    \:.
  \end{equation}
  From these results, we can deduce the variance:
  \begin{equation}
    \label{eq:MaxBoltzCan}
    \Var_\can(\N) = \frac{N}{4}
    \:.
  \end{equation}

\item Let us now consider the grand canonical ensemble in which the
  number $N$ of fermions fluctuates.  In this case, the expressions
  are simply obtained by averaging
  Eqs.~(\ref{eq:MeanNIMaxBoltzCan},\ref{eq:MeanNI2MaxBoltzCan}) over
  $N$:
  \begin{equation}
    \moyTg{\N} = \frac{\moyTg{N}}{2}
    \:,
  \end{equation}
  \begin{equation}
    \moyTg{(\N)^2} =
    \frac{\moyTg{N}}{4}
    + \frac{\moyTg{N^2}}{4}
    \:.
  \end{equation}
  From which we deduce:
  \begin{equation}
    \label{eq:VarNpMaxBoltzGC}
    \Var_\gc(\N) = 
    \frac{\moyTg{N}}{4}
    +\frac{1}{4} \: \Var_\gc(N)
    \:,
  \end{equation}
  where we have introduced the variance of the total number of
  particles $ \Var_\gc(N) = \moyTg{N^2} - (\moyTg{N})^2 $.
  The properties of this variance and its temperature dependence are
  discussed in Appendix~\ref{app:VarN}.

\end{itemize}
Let us remark that even though the relation between canonical and
grand canonical variances \eqref{eq:VarNpMaxBoltzGC} has been derived
in the classical Maxwell-Boltzmann regime it actually turns out to be
much more general, as we discuss in
Subsection~\ref{subsec:HeuristicRelationCanoGCano} below.

In the limit of high temperature $T \gg T_F$, the index variance
thus reads:
\begin{equation}
   \label{eq:VarNIAsympMaxBoltz}
  \left\lbrace
    \begin{array}{ll}
      \displaystyle
      \Var_{\can}(\N)  \simeq
      \frac{N}{4} & \text{ canonical,}
      \\[0.5cm]
      \displaystyle
      \Var_{\gc}(\N)  \simeq
      \frac{\moyTg{N}}{2} & \text{ grand canonical.}
    \end{array}
  \right.  
\end{equation}

\subsection{Canonical ensemble}
\label{sec:Can}

We have derived in the previous sections a general expression for
$\Var(\N)$, Eq.~(\ref{eq:VarIndexGen}), which involves the matrix
elements $A_{k,l}$ and occupation numbers. The coefficients $A_{k,l}$
are computed in Appendix~\ref{App:QuantAvrIndex}, and we studied the
occupation numbers in section~\ref{sec:OccNb}. In this section we will
combine these results to derive a more explicit expression for the
index variance of fermions in a harmonic trap $\Var_\can(\N)$, in
the canonical ensemble. In this case, since the total number $N$ of
fermions is fixed, $\Var_\can(N) = 0$ and
Eq.~(\ref{eq:VarIndexGen}) reduces to:
\begin{equation}
  \label{eq:VarIndexCan}
  \Var_\can(\N) =
  \frac{N}{4}
  - \underset{k\neq l}{\sum_{k,l=0}^\infty} \moyTc{n_k n_l} A_{k,l}^2
  \:.
\end{equation}
We will compute this variance first in the quantum regime $T \sim
T_Q$, then in the thermal regime $T \sim T_F$.

\subsubsection{Quantum regime}
\label{sec:QuantRegCan}

This regime can be reached for finite $\beta = 1/k_B T$ by letting the
number $N$ of fermions become large. Since we already know the
variance of $\N$ at zero temperature, Eq.~(\ref{eq:VarT0}), we
will focus on the difference between the variance at temperature $T$
and the one at zero temperature $T=0$:
\begin{equation}
  \Delta \Var_\can(\N) = 
  \Var_\can(\N) |_T - \Var_\can(\N) |_{T=0}
  \:.
\end{equation}
At $T=0$, Eq.~(\ref{eq:VarIndexCan}) is still valid, but with fixed
occupation numbers. Only the $N$ lowest energy levels are occupied,
thus
\begin{equation}
  n_k = \moyTc{n_k} = \left\lbrace
    \begin{array}{ll}
      1 & \text{if } k < N \:,\\
      0 & \text{if } k \geq N \:.
    \end{array}
  \right.
\end{equation}
Therefore, we have from Eq.~(\ref{eq:VarIndexCan}):
\begin{equation}
  \Delta \Var_\can(\N) =
   - \underset{k \neq l}{\sum_{k,l=0}^\infty} \moyTc{n_k n_l} A_{k,l}^2
  + \underset{k \neq l}{\sum_{k,l=0}^{N-1}} A_{k,l}^2
  \:.
\end{equation}
Since the main differences between the occupation numbers at zero and
finite temperature are visible near the Fermi level $N-1$, we shift
the indices in the sums to start the summation from the Fermi level:
\begin{align}
   \Delta \Var_\can(\N)
   = & \underset{k \neq l}{\sum_{k,l=0}^{N-1}}
   (1 - \moyTc{n_{N-k-1} n_{N-l-1}}) A_{N-k-1,N-l-1}^2
   - \underset{k \neq l}{\sum_{k,l=0}^{\infty}}
   \moyTc{n_{N+k} n_{N+l}} A_{N+k,N+l}^2
   \\
   & - 2 \sum_{k=0}^{N-1} \sum_{l=0}^\infty \moyTc{n_{N-k-1} n_{N+l}}
   A_{N-k-1,N+l}^2
   \:.
   \nonumber
\end{align}
First, for large $N$, we can let the summations go to infinity, as the
corrections are exponentially small with $N$. Then, since the
coefficients $A_{k,l}$ given by Eq.~(\ref{eq:Ant}) are non zero only
if $k$ and $l$ have different parity, we get:
\begin{align}
   \Delta \Var_\can(\N)
   = & \frac{1}{\pi^2} \underset{\neq \text{ parity}}{\sum_{k,l=0}^{\infty}}
   (1 - \moyTc{n_{N-k-1} n_{N-l-1}} - \moyTc{n_{N+k} n_{N+l}})
   \frac{1}{(k-l)^2}
   \\
   & - \frac{2}{\pi^2} \underset{\text{same parity}}{\sum_{k,l=0}^{\infty}}
   \moyTc{n_{N-k-1} n_{N+l}}
   \frac{1}{(k+l+1)^2}
   \:.
   \nonumber
\end{align}
We can then use Eq.~(\ref{eq:RelCovOccNbCan}) to evaluate the thermal
averages of products of occupation numbers. Introducing $k-l=2n-1$ in
the first sum, $k+l = 2n-2$ in the second one, and making use of
Eq.~(\ref{eq:SymOcNbFerm}) many cancellations occur, yielding a
compact expression:
\begin{equation}
  \label{eq:VarNbBoxQuantReg}
  \boxed{
  \Delta \Var_\can(\N)
  = \Fq(\beta\hbar \omega)
  = \frac{2}{\pi^2} \sum_{n=1}^\infty
  \frac{1}{2n-1}
  \frac{1}{\e^{\beta \hbar \omega(2n-1)} - 1}
  }
\end{equation}
involving a universal function $\Fq$, as argued below in
Section~\ref{sec:Universality}.  This is our final result for the
variance in the quantum regime for the canonical ensemble.
Remarkably, this formula for fermions involves Bose-Einstein factors,
like in the mean total energy~\cite{TexRou17} or its variance
\eqref{eq:VarTotalEnergyCanonical}.  This observation has a simple
origin: the system of fermions have particle-hole excitations of
bosonic nature. This well-known fact is as the heart of the
bosonization technique for 1D Fermi liquids (see~\cite{Hal81} for a
general reference and~\cite{SchMed96} for a discussion of bosonization
in the presence of a harmonic well).

From our result~(\ref{eq:VarNbBoxQuantReg}), we can extract the
asymptotic behaviours of the temperature dependent part of the variance:
\begin{equation}
  \label{eq:AsymptBoxQuantReg}
  \Delta \Var_\can(\N)
  \simeq
  \left\lbrace
    \begin{array}{ll}
      \displaystyle
      \frac{2}{\pi^2} \: \e^{-T_Q/T}
      &
      \text{ for } T \ll T_Q
      \\[0.4cm]
      \displaystyle
      \frac{T}{4 T_Q}
      &
      \text{ for } T \gg T_Q
      \:.
    \end{array}
  \right.
\end{equation}
The low temperature behaviour can be simply associated with the
existence of a gap $\hbar\omega$ in the excitation spectrum.

\subsubsection{Thermal regime}
\label{sec:ThermRegCan}

We now consider to the thermal regime, where the temperature is of the
order of the Fermi temperature $T \sim T_F$. We fix $\y = \beta N
\hbar \omega = T_F/T$ and let $N \to \infty$. In this case, we proceed
as in Subsection~\ref{sec:ThermRegKinE} where we analysed the potential
energy.  The study of the index is however more simple thanks to the
simplification mentioned in the beginning of
Section~\ref{sec:IndexFerm}, which makes the second term in
parenthesis in Eq.~\eqref{eq:92} vanish in the canonical ensemble.  As
a result, in this case it is sufficient to replace the occupation
numbers by the rescaled Fermi-Dirac distribution: $\moyTc{n_k} \simeq
f_\y(k/N) + \O(N^{-1})$ where $f_\y$ is given by Eq.~(\ref{eq:fyx}):
\begin{equation}
  \label{eq:VarIndexThermInterm}
  \Var_\can(\N) \simeq \frac{N}{4} - \sum_{k \neq l}
  f_\y  \left(\frac{k}{N} \right) f_\y  \left(\frac{l}{N} \right)
  (A_{k,l})^2
  \:.
\end{equation}
Let us first rewrite the double sum as:
\begin{equation}
  \sum_{k \neq l}
  f_\y  \left(\frac{k}{N} \right) f_\y  \left(\frac{l}{N} \right)
  (A_{k,l})^2
  = 2 \sum_{k=0}^\infty \sum_{p=1}^{k}
  f_\y \left(\frac{k}{N} \right) f_\y  \left(\frac{k-p}{N} \right)
  (A_{k,k-p})^2
  \:.
\end{equation}
Since $A_{k,l}$ is non zero only if $k$ and $l$ have different parity,
see Eq.~(\ref{eq:Ant}), the sum over $p$ involves only odd integers
$p=2n-1$. Replacing the summation over $k$ by an integral over $x=k/N$
gives:
\begin{equation}
  \label{eq:DblSumIndexTherm}
  \sum_{k \neq l}
  f_\y  \left(\frac{k}{N} \right) f_\y  \left(\frac{l}{N} \right)
  (A_{k,l})^2
  \simeq
  \frac{2N}{\pi^2}
  \int_0^\infty \sum_{n=1}^\infty \frac{f_\y(x)^2}{(2n-1)^2} \dd x
  = \frac{N}{4} \int_0^\infty f_y(x)^2 \dd x
  \:.
\end{equation}
Using this result in Eq.~(\ref{eq:VarIndexThermInterm}), along with
\begin{equation}
  N = \sum_{k=0}^\infty \moyTc{n_k} \simeq N \int_0^\infty f_\y(x)
  \dd x
  \:,
\end{equation}
yields
\begin{equation}
  \label{eq:LinkVarNpVarN}
  \Var_\can(\N) \simeq \frac{N}{4} \int_0^\infty  f_\y(x) (1- f_\y(x))
  \dd x
  = \frac{1}{4} \Var_\gc(N)
  \:,
\end{equation}
where $\Var_\gc(N)$ is the variance of the total number of particles
in the \textit{grand canonical} ensemble, where $\moyTg{N}$ must be
replaced by $N$. This non trivial relation between $\Var_\can(\N)$
and $\Var_\gc(N)$ relies on the specific properties of the matrix
elements $A_{k,l}$, thus on the nature of the observable $\N$. Using
the expression of this variance from Appendix~\ref{app:VarN}, we
obtain the final expression for the variance of the index in this
regime:
\begin{equation}
  \label{eq:VarBoxThermRegCan}
  \boxed{
    \Var_\can(\N)
    \simeq
    N \: \Ft \left(\y = \frac{T_F}{T} \right)
    = N \: \frac{1-\e^{-\y}}{4\y}
  }
\end{equation}
where the subleading $T=0$ contribution has been omitted. This is our
final result in the thermal regime for the canonical ensemble.  From
this general expression of the variance, we can extract its asymptotic
behaviours as function of the temperature:
\begin{equation}
  \label{eq:AsymptBoxThermReg}
   \Var_\can(\N) \simeq N \times
   \left\lbrace
     \begin{array}{cl}
       \displaystyle
       \frac{T}{4 T_F}
       &
       \text{ for } T \ll T_F
       \:,
       \\[0.5cm]
       \displaystyle
       \frac{1}{4}
       &
       \text{ for } T \gg T_F
       \:,
     \end{array}
     \right.
\end{equation}
First note that the high temperature limit $T \gg T_F$ matches with
the Maxwell-Boltzmann case, Eq.~(\ref{eq:MaxBoltzCan}), as it
should. In addition, the low temperature limit in this thermal regime,
$T \ll T_F$, smoothly matches the high temperature limit from the
quantum regime, $T \gg T_Q$, Eq.~(\ref{eq:AsymptBoxQuantReg}). This
indicates that there is no intermediate regime of temperature between
these two.  The low temperature result can be simply understood as the
classical result, $N/4$, reduced by the factor $T/T_F$ characteristic
of a degenerate Fermi gas~\cite{TexRou17}, as already
mentioned for the potential energy.

\subsection{Grand canonical ensemble}
\label{sec:GrandCan}

In the previous section we derived expressions for $\Var(\N)$ in
the canonical ensemble in both quantum and thermal regimes.  We now
perform a similar computation in the grand canonical ensemble. In this
case, the chemical potential $\mu$ is fixed, while the number of
fermions in the trap fluctuates. In order to easily compare the
results between the two ensembles, we will use the mean number of
particles $\moyTg{N}$ as a parameter instead of the chemical
potential. The two are related by Eq.~(\ref{eq:linkMuNgc}).

The mean occupation numbers $\moyTg{n_k}$ are given by the
Fermi-Dirac distribution~(\ref{eq:MeanOccupNbGC}). Occupations are
uncorrelated between different energy levels. This allows to rewrite
the general expression~(\ref{eq:VarIndexGen}) as
\begin{equation}
  \label{eq:VarNIGrdCanGen}
  \Var_\gc(\N) = \frac{\moyTg{N}}{4} + \frac{1}{4} \Var_\gc(N)
  - \underset{k\neq l}{\sum_{k,l=0}^\infty} \moyTg{n_k} \moyTg{n_l} A_{k,l}^2
  \:,
\end{equation}
where the variance of the total number of particles $\Var_\gc(N)$ is
studied in Appendix~\ref{app:VarN}. As before, we will first discuss the
quantum regime and then the thermal one.

\subsubsection{Quantum regime}
\label{sec:GrandCanQuantReg}

Again, we fix $\beta = 1/(k_B T)$ and let $\moyTg{N} \to \infty$. As
before, we focus on the difference
\begin{equation}
  \Delta \Var_\gc(\N) = 
  \Var_\gc(\N) |_T - \Var_\gc(\N) |_{T=0}
  \:.
\end{equation}
Using Eq.~(\ref{eq:VarNIGrdCanGen}), we can express this as
\begin{equation}
  \Delta \Var_\gc(\N) = 
  \frac{1}{4} \Var_\gc(N)
  - \underset{k \neq l}{\sum_{k,l=0}^\infty} \moyTg{n_k n_l} A_{k,l}^2
  + \underset{k \neq l}{\sum_{k,l=0}^{N-1}} A_{k,l}^2
  \:.
\end{equation}
We evaluated the same double sums in section~\ref{sec:QuantRegCan},
using only relations~(\ref{eq:RelCovOccNb}) and~(\ref{eq:SymOcNbFerm})
which hold in both ensembles. Therefore, our previous derivation is
still valid, and we have:
\begin{equation}
  - \underset{k \neq l}{\sum_{k,l=0}^\infty} \moyTg{n_k n_l} A_{k,l}^2
  + \underset{k \neq l}{\sum_{k,l=0}^{N-1}} A_{k,l}^2
  = \Delta \Var_\can(\N)
  \:.
\end{equation}
We obtain the final expression for the variance of the particle number:
\begin{equation}
  \label{eq:VarNbBoxQuantRegGC}
  \boxed{
    \Var_\gc(\N)
    = 
    \Var_\can(\N)
    + \frac{1}{4} \: \Var_\gc(N)
    =\left. \Var(\N) \right|_{T=0} +\Fq(\beta\hbar \omega)+\frac{1}{4} \: \Var_\gc(N)
  }
\end{equation}
where $\Fq(\xi)$ is given by Eq.~(\ref{eq:VarNbBoxQuantReg}). The variance in the grand canonical
ensemble is thus obtained from the canonical one by adding a term
proportional to the variance of the total number of particles. Using
the limiting behaviours of $\Var_\gc(N)$ given in
Appendix~\ref{app:VarN} along with Eq.~(\ref{eq:AsymptBoxQuantReg}),
we can straightforwardly deduce the asymptotic behaviours:
\begin{equation}
  \label{eq:AsymptBoxQuantRegGc}
  \Delta \Var_\gc(\N)
  \simeq
  \left\lbrace
    \begin{array}{ll}
      \displaystyle
      \frac{1}{2} \: \e^{-T_Q/2 T}
      &
      \text{ for } T \ll T_Q
      \:,
      \\[0.4cm]
      \displaystyle
      \frac{T}{2 T_Q}
      &
      \text{ for } T \gg T_Q
      \:.
    \end{array}
  \right.
\end{equation}
We again obtain an essential singularity at zero temperature, but
different from the canonical case ($\e^{-T_Q/T}$ \textit{vs}
$\e^{-T_Q/2T}$). This is due to the fact that in the grand-canonical
case, the leading contribution comes from the term proportional to the
total number of particles $\Var_\gc(N)$.

\subsubsection{Thermal regime}
\label{sec:GrandCanThermReg}

We now fix $\y = \moyTg{N} \beta \hbar \omega$ and let $\moyTg{N} \to
\infty$. As in the quantum regime, the last term in
Eq.~(\ref{eq:VarNIGrdCanGen}) was already computed in
section~\ref{sec:ThermRegCan}, and is given by
Eq.~(\ref{eq:DblSumIndexTherm}). Therefore, we straightforwardly
obtain:
\begin{equation}
  \label{eq:VarBoxThermRegGC}
  \boxed{
    \Var_\gc(\N) 
    =\Var_\can(\N) +  \frac{1}{4} \: \Var_\gc(N)
    \simeq
    \moyTg{N} \: \Ft\left(\y=\frac{T_F}{T}\right) + \frac{1}{4} \:
    \Var_\gc(N)
    }
\end{equation}
where $\Var_\can(\N)$ is here given by
Eq.~\eqref{eq:VarBoxThermRegCan} with $N$ substituted by $\moyTg{N}$.
In the r.h.s., we have neglected the subleading $T=0$ contribution,
Eq.~\eqref{eq:VarT0}. In this regime, the variance $\Var_\gc(N)$ can
be computed explicitly and is given by Eq.~(\ref{eq:VarNThermReg}):
\begin{equation}
  \Var_\gc(N) 
  \simeq \moyTg{N} \: \frac{1-\e^{-\y}}{\y}
  = 4 \: \moyTg{N} \: \Ft(\y)
  \:.
\end{equation}
Therefore, we can rewrite the index variance as
\begin{equation}
  \Var_\gc(\N) \simeq 2 \: \moyTg{N} \: \Ft(\y)
  = 2 \: \Var_\can(\N)
  \:.
\end{equation}
In this thermal regime, the index variance takes twice its canonical
value in the grand canonical case. We can thus straightforwardly
obtain its asymptotic behaviours as a function of $T$:
\begin{equation}
  \Var_\gc(\N)
  \simeq 
  \moyTg{N} \times
  \left\lbrace
    \begin{array}{ll}
      \displaystyle
      \frac{T}{2T_F}
      &
      \text{ for } T \ll T_F
      \:,
      \\[0.5cm]
      \displaystyle
      \frac{1}{2}
      &
      \text{ for } T \gg T_F
      \:,
    \end{array}
  \right.
\end{equation}
Again, we check that the low temperature limit $T \ll T_F$ smoothly
matches the limit $T \gg T_Q$ in the quantum regime,
Eq.~(\ref{eq:AsymptBoxQuantRegGc}). We also recover the
Maxwell-Boltzmann limit, Eq.~(\ref{eq:VarNIAsympMaxBoltz}) in the high
temperature limit, as expected, whereas the quantum regime again shows
the usual reduction factor $T/T_F$.


\subsection{Relation between the canonical and the grand canonical variances}
\label{subsec:HeuristicRelationCanoGCano}

We stress here that the relation between the variances in the
canonical and grand canonical ensembles is completely general: compare
\eqref{eq:MaxBoltzCan} and~\eqref{eq:VarNpMaxBoltzGC} in the
Maxwell-Boltzmann regime, or see Eq.~\eqref{eq:VarNbBoxQuantRegGC} in
the quantum regime and Eq.~\eqref{eq:VarBoxThermRegGC} in the thermal
regime. This relation can be recovered by a simple heuristic
argument: consider an extensive observable of the form $ A =
\sum_{i=1}^N a_i $. The average is also extensive, and we write it
under the form $\moyTg{A} = \moyTg{N} \: \moyTc{a}$.  We write the
variance $\Var_\gc(A) =\Var_\gc(N\,a)$ and assume that $a=A/N$ and $N$
are independent. As a result we get:
\begin{equation}
  \label{eq:VarObsCanGc}
  \Var_\gc(A) 
  = (\moyTg{N})^2 \: \Var(a) + (\moyTc{a})^2 \:  \Var_\gc(N)
  = \Var_\can(A) 
  + \bigg( \frac{\partial \moyTc{A}}{\partial N} \bigg)^2
  \Var_\gc(N)
  \:.
\end{equation}
Although this argument is not quite precise, in the case of the energy
$E$ of the system, it gives
\begin{equation}
  \Var_\gc(E)
  = \Var_\can(E) + \bigg( \frac{\partial \moyTc{E}}{\partial N} \bigg)^2
  \Var_\gc(N)
  \:,
\end{equation}
which is a well-known relation in statistical
mechanics~\cite{PatBea11,TexRou17}. Applied to the case of $\N$,
Eq.~(\ref{eq:VarObsCanGc}) becomes:
\begin{equation}
  \Var_\gc(\N) = \Var_\can(\N) + \frac{1}{4} \: \Var_{\gc}(N)
  \:,
\end{equation}
which is clearly the relation obtained several times above~(\ref{eq:VarNpMaxBoltzGC},\ref{eq:VarNbBoxQuantRegGC},\ref{eq:VarBoxThermRegGC}).


\subsection{Universality}
\label{sec:Universality}

All our discussion so far has focused on the example of the harmonic
trap as in this case the one body wave-functions are known, which
makes the computations more explicit.  Some of our results can however
be extended to any type of confining potential, assuming a single
minimum for simplicity.

Let us first discuss the zero temperature result, $\left. \Var(\N)
\right|_{T=0}$, given by Eq.~(\ref{eq:VarT0}) for the harmonic trap. A
similar result has been obtained for a system of fermions confined in
an infinite square well. In this case, the variance of the number of
particles in the right half part of the trap can be found in Eq.~(40)
of Ref.~\cite{CalMinVic12}:
\begin{equation}
  \label{eq:VarT0Box}
  \left. \Var(\N) \right|_{T=0}^\text{Box} \simeq 
  \frac{1}{2 \pi^2} \ln N
  + \frac{1+\gamma+2\ln 2}{2\pi^2}
  \:.
\end{equation}
The variations of $\N$ correspond to particles crossing the origin,
thus $\left. \Var(\N) \right|_{T=0}$ measures the quantum fluctuations
through the origin. The leading log-terms in (\ref{eq:VarT0}) and
(\ref{eq:VarT0Box}) coincide, and thus is a bulk property (which can
be related to~(\ref{eq:BulkCorrelFct})).  However,the subleading
constants are different. We thus conclude that they are sensitive to
the precise form of the potential. Therefore, we expect only the
leading log-term to be universal.

At finite temperature, the fluctuations are controlled by the two
scaling functions $\Fq$ and $\Ft$, for the quantum and thermal regimes
respectively. These functions are determined by the occupation numbers
$\moyT{n_k}$, which depend on the spectrum $\{ \varepsilon_n \}$ and
thus on the potential, and the matrix elements $A_{k,l}$. These latter
depend only on the values of the wave functions $\psi_n$ at the center
of the trap, see Eq.~(\ref{eq:AklCenterTrap}). These can thus be
derived by semiclassical methods, such as the WKB approximation, for
any kind of potential and exhibit universal properties. In particular,
one can show that the expression~(\ref{eq:Ant})
of these matrix elements for large quantum numbers is universal.

In the quantum regime, the derivation of
Sections~\ref{sec:QuantRegCan} and \ref{sec:GrandCanQuantReg} is based
on this universal expression of the matrix elements $A_{k,l}$ and on
two properties of the occupation numbers,
Eqs.~(\ref{eq:RelCovOccNbCan}) and (\ref{eq:SymOcNbFerm}). The first
relation is universal, as discussed in Section~\ref{sec:OccNb}. The
second property~(\ref{eq:SymOcNbFerm}), which implies the symmetry of
the occupation numbers around the Fermi level, holds only for a linear
spectrum. However, any regular spectrum $\{ \varepsilon_n \}$ can be
linearised near the Fermi level and thus~(\ref{eq:SymOcNbFerm}) is
also universal in the vicinity of the Fermi level. The temperature
scale $T_Q$ is then defined from the gap at the Fermi level:
\begin{equation}
  \label{eq:DefTq}
  T_Q = \frac{\varepsilon_{N}-\varepsilon_{N-1}}{k_B}
  \:.
\end{equation}
Therefore, the derivation of Sections~\ref{sec:QuantRegCan} and
\ref{sec:GrandCanQuantReg} can be extended to any confining potential,
and the scaling function $\Fq$ is thus universal, and given by
expression~(\ref{eq:FqUniv}).

In the thermal regime,
controlled by the scale
\begin{equation}
  T_F = \frac{\varepsilon_N+\varepsilon_{N-1}}{2k_B}
  \:,
\end{equation}
all the spectrum contributes to the variance $\Var(\N)$. Therefore, we
do not expect the function $\Ft$ to be universal. This can be shown
explicitly from the relation~(\ref{eq:LinkVarNpVarN}) which states
that this scaling function is proportional to the variance of the
total particle number in the grand canonical ensemble: $\Ft(T_F/T) =
\Var_\gc(N)/4$. We do not know this function
  explicitly, however we show in Appendix~\ref{app:VarN} that its
  limiting behaviours are only controlled by the exponent governing
  the one body density of states $\rho(\varepsilon) \propto
  \varepsilon^{\alpha-1}$:
\begin{equation}
  \Ft \left( \y = \frac{T_F}{T} \right)
  \simeq
  \frac{1}{4}
  \left\lbrace
    \begin{array}{cl}
      1 & \text{for } T \gg T_F
      \:,
      \\[0.2cm]
      \displaystyle
      \alpha \frac{T}{T_F} & \text{for } T \ll T_F
      \:,
    \end{array}
  \right.
\end{equation}
which thus depends explicitly on $\alpha$.


\subsection{Numerical simulations}

We now compare our results with numerical simulations.  Generating
numerically realisations of the positions of the fermions is quite
difficult. But in the grand canonical ensemble, we can use the fact
that the positions of the fermions form a determinantal point
process~\cite{Joh07,DeLDoMajSch15,DeLDoMajSch16}, with kernel:
\begin{equation}
  K_T(x,x') = \sum_k \moyTg{n_k} \psi_k(x) \psi_k(x')
  \:,
\end{equation}
where $\psi_k$ are the one-particle wave functions of the harmonic
oscillator~(\ref{eq:WaveFctHarmOsc}).  We reproduce in
Appendix~\ref{app:Numerics} an algorithm described in
Refs.~\cite{HoiKriPerVir06,LavMolRub15} which allows to sample such
point processes. Therefore, we performed simulations only in the
grand canonical case.

\begin{figure}[!ht]
  \centering
  \includegraphics[width=0.6\textwidth]{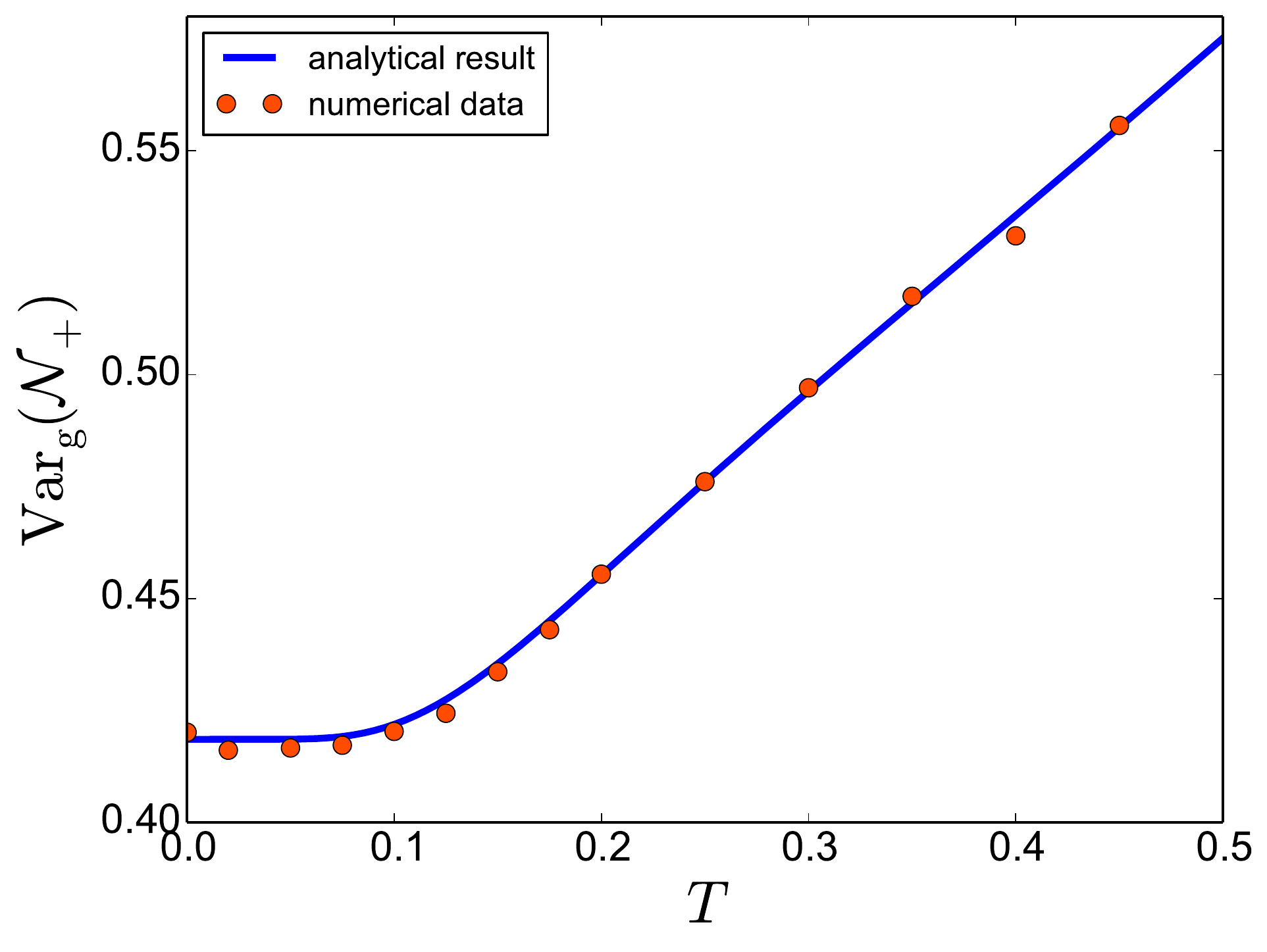}
  \caption{\it Variance of the index number $\N$, in the grand
    canonical ensemble for $\moyTg{N}=100$ as a function of
    temperature (we set $\hbar = \omega = m = k_B = 1$). The points
    are obtained by averaging over $10^5$ realisations, while the
    solid line corresponds to our analytical result,
    Eq.~(\ref{eq:FullVarGC}).}
  \label{fig:NumBox}
\end{figure}

We generate realisations of the positions of fermions and compute the
corresponding value of $\N$ for each of them. From the set of data, we
compute numerically $\Var(\N)$. We studied the quantum regime, with
$\moyTg{N}=100$. We did not investigate the thermal regime because the
computational efforts increase with the temperature, as more energy
levels contribute. This makes it difficult to obtain enough statistics
in the thermal regime where $T = \O(\moyTg{N})$. Our result in the
grand canonical ensemble~(\ref{eq:VarNbBoxQuantRegGC}), combined with
the previously known zero temperature expression~(\ref{eq:VarT0})
read:
\begin{equation}
  \label{eq:FullVarGC}
  \Var_\gc(\N) \simeq
   \frac{1}{4} \: \Var_\gc(N)
    + \Fq(\beta \hbar \omega)
    + \frac{1}{2\pi^2} \ln \moyTg{N} + c
  \:,
\end{equation}
where $\Fq(\beta \hbar \omega)$ is given by
Eq.~(\ref{eq:VarNbBoxQuantReg}) and $c$ is a constant,
see Eq.~(\ref{eq:VarT0}). We used this expression to compare the
numerical data to our analytical result. They show an excellent
agreement, see Fig.~\ref{fig:NumBox}.

\section{Conclusion}
\label{sec:Conclusion}

In this paper, we have introduced a general method to study certain
many body observables of the form of sums of one body observables of
the positions of fermions in a confining trap.  We have applied our
method to the study of the number $\N$ of fermions on the positive
axis, in the case of a harmonic well. We have obtained explicit
expressions for the variance of these observables, both in the quantum
regime $T \sim T_Q = \hbar \omega/{k_B}$ and the thermal regime $T
\sim T_F = N \hbar \omega/{k_B}$ in terms of two scaling functions,
one universal and the other not. We have shown that these expressions
smoothly match. We note that the fluctuations of linear statistics
were recently studied by Johansson and Lambert in the grand canonical
ensemble~\cite{JohLam15}. They considered the case where each of the
fermions positions $x_i$'s are scaled like $N^{-\delta}$ while the
temperature scales like $T\sim N^{\alpha}$. The results presented in
this paper thus corresponds to $\delta=0$ (which in their terminology
corresponds to ``macroscopic scale''). In addition, here, we have
studied the crossover between the quantum regime ($\alpha=0$ in the
notation of Ref.~\cite{JohLam15}) and the thermal regime ($\alpha=1$).

We have emphasised the difference between
the statistical ensemble by computing the variance of $\N$ both in the
canonical and grand-canonical ensembles. This difference is due to
the fact that we have considered the fluctuations of a \textit{global}
observable, $\mathcal{N}_+=\sum_n\Theta(x_n)$.  On the other,
\textit{local} quantities, such as the two point density-density
correlation functions, are ensemble independent~\cite{DeLDoMajSch16}.

The computation of these fluctuations in the microcanonical ensemble
is still an open question. Indeed, our derivation made extensive use
of relation~(\ref{eq:RelCovOccNb}) which no longer holds in the
microcanonical case. We have studied the first moments of the
distribution of the observable $\N$ at finite temperature. Computing
the full distribution of $\N$ would be an interesting but much more
challenging question.


\section*{Acknowledgements}

This research was supported by ANR grant ANR-17-CE30-0027-01 RaMaTraF.
We acknowledge stimulating discussions with Olivier Giraud.


\appendix


\section{Variance of the total particle number in the grand canonical
  ensemble for any confining potential}
\label{app:VarN}

As discussed in Section~\ref{sec:OccNb}, in the grand-canonical
ensemble the mean occupation numbers $\moyTg{n_k}$ are given by the
Fermi-Dirac distribution, Eq.~(\ref{eq:MeanOccupNbGC}). In addition,
occupations are uncorrelated: $\moyTg{n_k n_l} = \moyTg{n_k} \:
\moyTg{n_l}$. This allows to easily evaluate the first moments of the
fluctuating total number of particles $\Ng = \sum_k n_k$:
\begin{equation}
  \label{eq:MoyNgApp}
  \moyTg{\Ng} = \sum_k \moyTg{n_k}
  = \sum_{k} \frac{1}{\e^{\beta(\varepsilon_k-\mu)}+1}
  \:.
\end{equation}
This first relation links the mean number of particles $\moyTg{\Ng}$
to the chemical potential $\mu$. The variance reads:
\begin{equation}
  \Var_\gc(\Ng) = \sum_k \Var_\gc(n_k)
  + \sum_{k \neq l} \underbrace{\Cov_\gc(n_k, n_l)}_{=0}
  = \sum_k \moyTg{n_k} (1-\moyTg{n_k})
  \:.
\end{equation}
Therefore, we obtain:
\begin{equation}
  \label{eq:VarNgApp}
  \Var_\gc(\Ng)
  = \sum_k \frac{1}{\e^{\beta(\varepsilon_k-\mu)}+1}
  \left(
    1 - \frac{1}{\e^{\beta(\varepsilon_k-\mu)}+1}
  \right)
  \:.
\end{equation}
We can use these expressions to study the limiting behaviours of this
variance for any confining potential of the form $V(x) \propto |x|^p$.
The spectrum can be determined from a WKB approximation
\begin{equation}
  \int_{-x_t}^{x_t} \sqrt{2m(\varepsilon_n - V(x))} = 
  \left( n + \frac{1}{2} \right) \pi \hbar
  \:,
\end{equation}
where $x_t$ is the turning point $V(x_t) = \varepsilon_n$. This
condition gives the scaling
\begin{equation}
  \varepsilon_n \sim n^{1/\alpha}
  \:,
  \quad
  \alpha = \frac{1}{2} + \frac{1}{p}
  \:.
\end{equation}
We can check that $\alpha = 1$ for the harmonic potential while
$\alpha=1/2$ corresponds to the infinite square well. In the
continuum limit, this spectrum gives a density of states of the form
\begin{equation}
  \rho(\varepsilon) = A \frac{\varepsilon^{\alpha - 1}}{\delta^\alpha}
  \:,
\end{equation}
where $\delta$ is an energy scale and $A$ is a dimensionless
parameter. For a trap containing on average $\moyTg{N}$ particles, the
Fermi energy is given by $\varepsilon_F = (\alpha
\moyTg{N}/A)^{1/\alpha} \delta$. Therefore, we define the Fermi
temperature as
\begin{equation}
  T_F = \frac{\varepsilon_F}{k_B} = 
  \left( \frac{\alpha \moyTg{N}}{A} \right)^{1/\alpha}
  \frac{\delta}{k_B}
  \:.
\end{equation}
The temperature scale $T_Q$ can be defined from the gap at the Fermi
level~(\ref{eq:DefTq}), which gives
\begin{equation}
  T_Q = \frac{1}{k_B \rho(\varepsilon_F)} = \frac{1}{\alpha
    \moyTg{N}} \: T_F
  \:.
\end{equation}

In the quantum regime, the spectrum can be linearised near the Fermi
level:
\begin{equation}
  \varepsilon_{\moyTg{N}+n} \simeq \varepsilon_F + n \: k_B T_Q
  \:.
\end{equation}
The variance $\Var_\gc(N)$ is thus universal in this regime. In the
low temperature limit $T \ll T_Q$, Eq.~(\ref{eq:MoyNgApp}) imposes
that the chemical potential is fixed to the middle of the gap $\mu
\simeq \varepsilon_F + k_B T_Q/2$. Using this value in
Eq.~(\ref{eq:VarNgApp}), we can study the low temperature limit of the
variance of $N$. The leading contribution comes from the two levels
$\varepsilon_{\moyTg{N}-1}$ and $\varepsilon_{\moyTg{N}}$ which are
the closest to the chemical potential $\mu$. This gives
\begin{equation}
  \Var_\gc(\Ng) \simeq 2 \: \e^{-T_Q/2T}
  \:,
  \quad
  T \ll T_Q
  \:.
\end{equation}

\begin{figure}[!ht]
  \centering
  \includegraphics[width=0.45\textwidth]{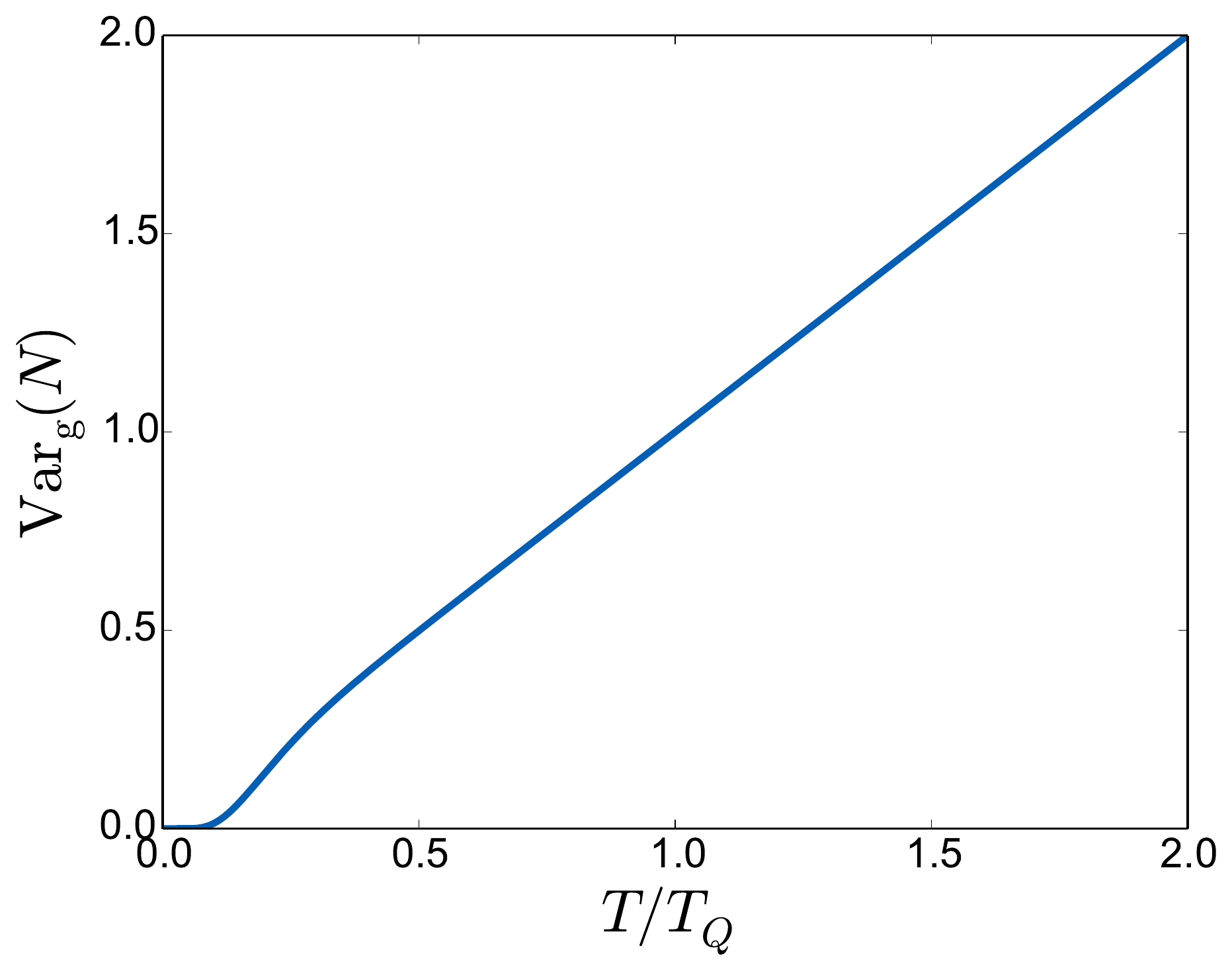}
  \includegraphics[width=0.45\textwidth]{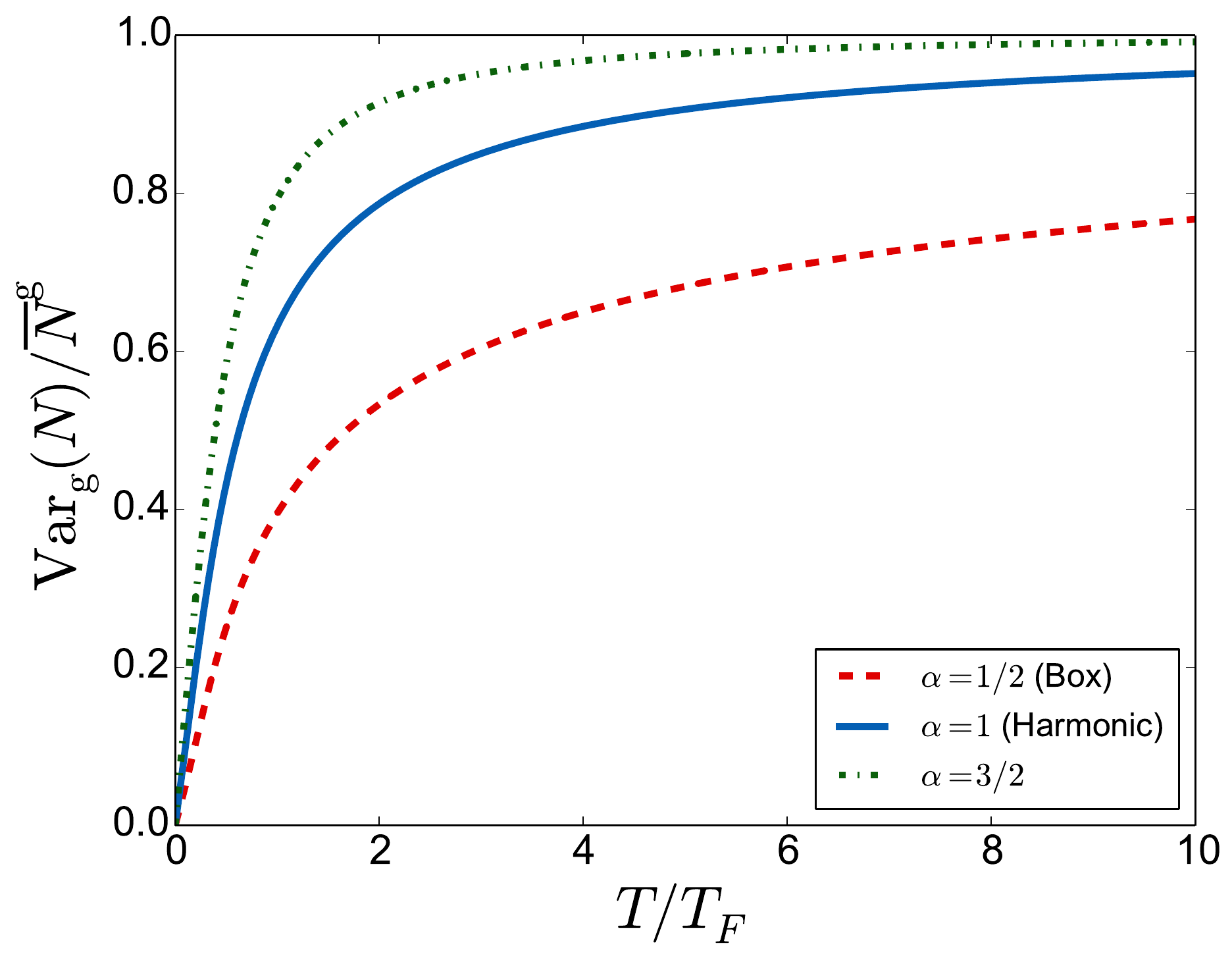}
  \caption{
    \textit{Variance of the total number $\Ng$ of fermions in the grand
    canonical ensemble, as a function of the temperature $T$. }
    Left:
    \textit{quantum regime $T \sim T_Q$ obtained
    from~(\ref{eq:VarNgApp}). This function is universal. }
    Right: \textit{thermal regime $T \sim T_F$,
    given by Eq.~(\ref{eq:VarNThermReg}) for different types of
    confinment: $\alpha = 1$ (harmonic), $\alpha=1/2$ (infinite well)
    and $\alpha = 3/2$.}}
  \label{fig:PltVarN}
\end{figure}

In the thermal regime $T \sim T_F$, the sums can be replaced by
integrals over the energy $\varepsilon$. Eq.~(\ref{eq:MoyNgApp})
becomes:
\begin{equation}
  \moyTg{N} = \sum_k  \moyTg{n_k} \simeq
  \int_0^\infty \frac{\rho(\varepsilon) \: \dd
    \varepsilon}{\e^{\beta(\varepsilon - \mu)} + 1}
  = - \frac{A}{(\beta \delta)^\alpha}  \Gamma(\alpha) \:
  \mathrm{Li}_\alpha(- \e^{\beta \mu})
  \:,
\end{equation}
where we recall the $\mathrm{Li}_\alpha(z) = \sum_{k=1}^\infty
z^k/k^\alpha$. This last relation is more conveniently expressed in terms of 
dimensionless variables:
\begin{equation}
  \label{eq:MuTilde}
  - \y^{-\alpha} \: \Gamma(\alpha+1) \:
  \mathrm{Li}_\alpha(- \e^{\y \tilde{\mu}}) = 1
  \:,
  \quad
  \y = \frac{T_F}{T}
  \text{ and } 
  \tilde{\mu} = \frac{\mu}{\varepsilon_F}
  \:.
\end{equation}
This relation fixes the rescaled chemical potential $\tilde{\mu}$ in
terms of the rescaled inverse temperature $\y$. A similar computation
for the variance~(\ref{eq:VarNgApp}) gives:
\begin{equation}
  \label{eq:VarNThermReg}
  \Var_\gc(\Ng) = \frac{1}{\beta} \frac{\partial \moyTg{N}}{\partial \mu}
  = - \moyTg{N} \: \y^{-\alpha} \: \Gamma(\alpha+1)
  \mathrm{Li}_{\alpha-1}(- \e^{\y \tilde{\mu}})
  \:.
\end{equation}
These two relations~(\ref{eq:MuTilde}) and~(\ref{eq:VarNThermReg})
allow to plot this variance for different confining potentials,
corresponding to different values of $\alpha$, see
Fig.~\ref{fig:PltVarN}, right. In the high-temperature limit, the
variance reaches the classical limit
\begin{equation}
  \Var_\gc(\Ng) \simeq \moyTg{N}
  \:,
  \quad
  T \gg T_F
  \:.
\end{equation}
The other limiting case is:
\begin{equation}
  \label{eq:VarNLimitIntermReg}
  \Var_\gc(\Ng) \simeq \alpha \moyTg{N} \frac{T}{T_F} = \frac{T}{T_Q}
  \:,
  \quad
  T_Q \ll T \ll T_F
  \:,
\end{equation}
which shows once again the reduction factor $T/T_F$, as usual in the
degenerate Fermi gas~\cite{TexRou17}. See Fig.~\ref{fig:PltVarN} for
plots of this variance $\Var_\gc(\Ng)$ in both the quantum and thermal
regimes, for different confining potentials. In the case of the
harmonic trap ($\alpha=1$),
Eqs.~(\ref{eq:MuTilde},\ref{eq:VarNThermReg}) reduce to the simple
expression
\begin{equation}
  \Var_\gc(N) = \frac{1-\e^{-\y}}{\y}
  \:,
  \quad
  \y = \frac{T_F}{T}
  \:.
\end{equation}

\section{Saddle point estimate}
\label{app:SaddlePt}

Let us consider integrals of the form:
\begin{equation}
  I(N) = \int_\mathcal{C} \dd z \: g(z) \: \e^{-N \phi(z)} \:,
\end{equation}
where $\mathcal{C}$ is any contour in the complex plane, $g$ and
$\phi$ are any given smooth functions. We want to estimate this
integral in the limit of large $N$ by using a saddle point method. The
saddle point $z_\star$ is given by
\begin{equation}
  \phi'(z_\star) = 0 \:.
\end{equation}
Let us make the change of variable $z = z_\star + t/\sqrt{N}$ and
deform the contour $\mathcal{C}$ such that $t$ is real. We can
expand $\phi$ and $g$ near $z_\star$:
\begin{align}
  \phi(z) &= \phi(z_\star) + \frac{t^2}{2N} \phi''(z_\star) 
  + \frac{t^3}{6N^{3/2}} \phi^{(3)}(z_\star)
  + \frac{t^4}{24 N^2} \phi^{(4)} +\O(N^{-5/2}) \:,\\
  g(z) &= g(z_\star) + \frac{t}{\sqrt{N}} g'(z_\star)
  + \frac{t^2}{2N} g''(z_\star) + \O(N^{-3/2}) \:.
\end{align}
Using these expansions, the integral $I(N)$ becomes:
\begin{align}
  \nonumber
  I(N) &= \frac{\e^{-N \phi(z_\star)}}{\sqrt{N}} \int_\mathbb{R} \dd t \:
  \e^{-t^2 \phi''(z_\star)/2} \:
  \\
  \nonumber
  &\times \left(
    1 - \frac{t^3}{6 \sqrt{N}} \phi^{(3)}(z_\star)
    - \frac{t^4}{24N} \phi^{(4)}(z_\star)
    + \frac{t^6}{72N} (\phi^{(3)}(z_\star))^2 + \O(N^{-3/2})
  \right)
  \\
  &\times \left(
    g(z_\star) + \frac{t}{\sqrt{N}} g'(z_\star)
    + \frac{t^2}{2N} g''(z_\star) + \O(N^{-3/2})
  \right) \:.
\end{align}
Computing the Gaussian integrals yields:
\begin{align}
  \nonumber
  I(N) = \e^{-N \phi(z_\star)}
  \sqrt{\frac{2\pi}{N \phi^{(2)}(z_\star)}}
  &
  \left[
    g(z_\star)
    + \frac{1}{N} \left(
      \frac{g^{(2)}(z_\star)}{2 \phi^{(2)}(z_\star)}
      - \frac{g(z_\star)\phi^{(4)}(z_\star)}{8(\phi^{(2)}(z_\star))^2}
    \right. \right.
  \\
  &\left. \left.
      + \frac{5 g(z_\star) (\phi^{(3)}(z_\star))^2}{24 (\phi^{(2)}(z_\star))^3}
      - \frac{g'(z_\star) \phi^{(3)}(z_\star)}{2 (\phi^{(2)}(z_\star))^2}
    \right)
    +\O(N^{-2})
  \right]
  \:.
\end{align}
We used this expression in section~\ref{subsec:SaddlePt} to obtain the
$\O(N^{-1})$ corrections to the covariance $\moyTc{n_k n_l} -
\moyTc{n_k} \moyTc{n_l}$.

\section{Matrix elements}
\label{App:QuantAvrIndex}

In this section we compute the coefficients $A_{k,l}$ associated to
the index variance, Eq.~(\ref{eq:AklHarmOsc}), which are obtained from
the quantum average. Since the index $\N$ is a linear
statistics~(\ref{eq:defLinStat}) for $\ls(x) = \ls(x)^2 = \Theta(x)$, we can
straightforwardly obtain the ``diagonal terms''
\begin{equation}
  B_k = A_{k,k} = \int_0^\infty \psi_k(x)^2 \dd x 
  = \frac{1}{2}
  \:.
\end{equation}
The ``off-diagonal'' coefficients are given by:
\begin{equation}
  A_{k,l} =
  \int_{0}^\infty \psi_{k}(x) \psi_{l}(x) \dd x
  \:.
\end{equation}
This integral can be
computed analytically. Indeed, let us compute the derivative of
$\psi_k(x) \psi_l'(x) - \psi_k'(x) \psi_l(x)$:
\begin{equation}
  \frac{\dd}{\dd x} \left( \psi_k(x) \psi_l'(x) - \psi_k'(x)
\psi_l(x) \right)
= \psi_k(x) \psi_l''(x) - \psi_k''(x) \psi_l(x)
\end{equation}
since the other terms cancel out. Using now that $\psi_k$ is
solution of the Schr\"odinger equation $-\frac{\hbar^2}{2m}
\psi_k''+V \psi_k = \varepsilon_k \psi_k$, we get:
\begin{equation}
  \frac{\dd}{\dd x} \left( \psi_k(x) \psi_l'(x) - \psi_k'(x)
\psi_l(x) \right)
= \frac{2m}{\hbar^2}(\varepsilon_k-\varepsilon_l)  \psi_k(x)
\psi_l(x)
\:.
\end{equation}
This relation allows us to directly compute the integral
\begin{equation}
  \label{eq:AklCenterTrap}
  A_{k,l} = \int_{0}^{\infty} \psi_k(x)
  \psi_l(x) \: \dd x
  = -\frac{\hbar^2}{m} \frac{1}{\varepsilon_k - \varepsilon_l}
  \left(
   \psi_k(0) \psi_l'(0) - \psi_k'(0) \psi_l(0)
   \right)
   \:.
\end{equation}
Using the expression of the wave functions,
Eq.~(\ref{eq:WaveFctHarmOsc}) and properties of the Hermite
polynomials~\cite{GraRyz94}, we obtain that $A_{k,l}$ is zero if $k$
and $l$ have the same parity, and
\begin{equation}
  A_{2m,2n+1} = \frac{
    (-2)^{n+m+1} \Gamma(m+\frac{1}{2}) \Gamma(n+\frac{3}{2})
  }{
    \pi \sqrt{2\pi} \sqrt{(2n+1)!(2m)!} (2m-2n-1)
  }
  \:.
\end{equation}
Since we only need the square of these coefficients, we can simplify
this expression using properties of the $\Gamma$ function:
\begin{equation}
  (A_{2m,2n+1})^2 = \frac{\Gamma(m+\frac{1}{2}) \Gamma(n+\frac{3}{2})}
  {\pi^2 n! m!} \frac{1}{(2m-2n-1)^2}
  \:.
\end{equation}
We are interested in the limit in which the number $N$ of fermions is
large. We expect the variance of $\N$ to be dominated by the
fluctuations near the Fermi level. Therefore, it is enough to estimate
the coefficients $A_{k,l}$ for $k$ and $l$ of order $N$. Therefore,
for $k = Nt$ and $l = k + p$, we get:
\begin{equation}
  \label{eq:Ant}
    (A_{Nt,Nt+p})^2 \simeq \left\lbrace
      \begin{array}{ll}
        0 & \text{ if } p \text{ is even,} 
        \\[0.2cm]
        \displaystyle
        \frac{1}{p^2 \pi^2}
        & \text{ if } p \text{ is odd.}
      \end{array}
    \right.
\end{equation}
Note that these are the leading contributions to the coefficients
$A_{k,l}$: they also receive $\O(N^{-1})$ corrections. In addition, we
have only considered the case where $k,l = \O(N)$ with $k-l=\O(1)$. It
is clear from the final expression of $A_{k,l}$~(\ref{eq:Ant}) that
the case $k-l = \O(N)$ gives only subleading corrections.


\section{Numerical simulations of determinantal point processes}
\label{app:Numerics}

A determinantal point process is a random point process $\{ x_n \}$
which is entirely characterised by a \textit{kernel} $K(x,y)$. All
$n$-points correlations functions can be expressed as $n\times n$
determinants involving the kernel $K$, see Eq.~(\ref{eq:nPtCorrFct}).
We consider such a process with a kernel
\begin{equation}
  \label{eq:KernProc}
  K(x,y) = \sum_{k=0}^\infty \lambda_k \psi_k^*(x) \psi_k(y)
  \:,
\end{equation}
where $0 \leq \lambda_k \leq 1$ and
\begin{equation}
  \int_{-\infty}^\infty \psi_k^*(x) \psi_l(x) \dd x = \delta_{k,l}
  \:.
\end{equation}
A general method to generate numerically realisations of this process
was introduced in Ref.~\cite{HoiKriPerVir06}. We reproduce here a
similar algorithm described in~\cite{LavMolRub15}. This algorithm was
also used in the physics literature, see
e.g. Refs.~\cite{ScaZacTor09,TorScaZac08}. It relies on the following
theorem: introduce a set of Bernoulli random variables $n_k = 0$ or
$1$, with mean value $\overline{n_k} = \lambda_k$. Then, the
determinantal point process with kernel
\begin{equation}
  \tilde{K}(x,y) = \sum_{k=0}^\infty n_k \psi_k^*(x) \psi_k(y)
\end{equation}
has the same statistics as the original process with
kernel~(\ref{eq:KernProc}). In terms of fermions, this means that
picking a realisation of the positions of the particles in the grand
canonical ensemble is equivalent to first picking a quantum state $\{
n_k \}$ from the Gibbs measure~(\ref{eq:MeasureGC}) and then
generating a realisation of the positions from that quantum state.
Using this property, one can generate realisations of the
determinantal point process using the following procedure:
\begin{enumerate}
\item Generate the index $M$ of the highest occupied level, using that
  \begin{equation}
    \mathrm{Proba}(M=m) = \lambda_m \prod_{i>m} (1-\lambda_i)
    \:.
  \end{equation}
\item Generate the occupation numbers for $k < M$, from the measure
  \begin{equation}
    \mathrm{Proba}(n_k = 1) = \lambda_k
    \:.
  \end{equation}
  Set $n_M = 1$ and $n_p = 0$ for $p > M$. Note that this realisation
  will contain $N = \sum_k n_k$ points. Denote also $\{ k_n
  \}_{n=1,\ldots,N}$ the indices of the occupied levels ($n_{k_i} =
  1$) and $\vec{v}(x) = (\psi_{k_1}(x), \ldots, \psi_{k_N}(x))^T$.
\item Pick the first point $X_1$ from the distribution
  \begin{equation}
    p_1(x) = \frac{1}{N} \sum_{p=1}^{N} \abs{\psi_{k_p}(x)}^2
    = \frac{||\vec{v}(x)||^2}{N}
    \:,
  \end{equation}
  and introduce $\vec{e}_1 = \vec{v}(X_1)/||\vec{v}(X_1)||$.
\item Knowing the positions $\{ X_1, \ldots , X_{n} \}$ of the first $n$
  points and the set of unit vectors $(\vec{e}_1,\ldots,\vec{e}_{n})$
  generate the position $X_{n+1}$ of the next point from the
  distribution
  \begin{equation}
    p_{n+1}(x) = \frac{1}{N-n} \left(
      ||\vec{v}(x)||^2
      - \sum_{j=1}^{n} \abs{\vec{e}_j{}^* \cdot \vec{v}(x)}^2
    \right)
    \:.
  \end{equation}
  Construct $\vec{e}_{n+1} = \vec{w}_{n+1} / ||\vec{w}_{n+1}||$, where
  \begin{equation}
    w_{n+1} = \vec{v}(X_{n+1})
    - \sum_{j=1}^{n} (\vec{e}_j{}^* \cdot \vec{v}(X_{n+1}))
    \: \vec{e}_j
    \:.
  \end{equation}
\end{enumerate}
This procedure gives a realisation $(X_1, \ldots, X_N)$ of the
determinantal point process with
kernel~(\ref{eq:KernProc}). Generating the points from the rather complex
distributions $p_n(x)$ can be done using rejection sampling~\cite{Dev86}.



\end{document}